\documentclass[sn-basic]{sn-jnl}

\usepackage{graphicx}%
\usepackage{multirow}%
\usepackage{amsmath,amssymb,amsfonts}%
\usepackage{amsthm}%
\usepackage{mathrsfs}%
\usepackage[title]{appendix}%
\usepackage{xcolor}%
\usepackage{textcomp}%
\usepackage{manyfoot}%
\usepackage{booktabs}%
\usepackage{algorithm}%
\usepackage{algorithmicx}%
\usepackage{algpseudocode}%
\usepackage{listings}%

\begin{document}

\title[Numerical model of fast electron energy deposition]{Numerical model of fast electron energy deposition in interstellar molecular gas}

\author*[1]{\fnm{Aleksandr} \sur{Nesterenok}}\email{nesterenokastro@gmail.com}

\affil*[1]{\orgname{Ioffe Institute}, \orgaddress{\street{Polytechnicheskay st, 26}, \city{Saint Petersburg}, \postcode{194021}, \country{Russian Federation}}}

\abstract{The energy deposition of fast electrons in interstellar molecular gas is considered. We use the rotationally resolved cross sections for electron-impact excitation of H$_2$ molecule that were calculated using the adiabatic-nuclei molecular convergent close-coupling method. The initial electron energy distribution is assumed mono-energetic, and the differential equation for electron energy distribution is solved. We compare calculated energy deposition parameters with the results of similar studies in which the Monte Carlo approach was used. It is shown that about 11 per cent of the initial energy of fast electrons goes into direct ro-vibrational excitation of energy levels of H$_2$ molecule including pure rotational excitation in neutral molecular gas. About 7 per cent of initial electron energy goes into the excitation to $v=1$ vibrational state of H$_2$ molecule, most of this energy eventually converts into emission of transitions at near-infrared wavelengths. For ro-vibrational levels with $v \geq 3$, the electron-impact excitation to electronic states followed by downward radiative transitions to the ground electronic state is the dominant mechanism of excitation. The yields for excitation to vibrational states via radiative cascading from excited electronic states are found to be $1.5-2$ times higher than were obtained in previous studies.}


\maketitle

%
\section{Introduction}\label{sec1}
The energetic electrons are produced in the astrophysical environment as a result of interactions of high-energy particles with interstellar gas clouds, or as a result of photoionisation and Compton ionisation of gas species in the vicinity of sources of intense X-ray and gamma-ray radiation. The electrons deposit their energy through the ionisation and excitation of atoms and molecules. This changes the gas ionisation fraction and leads to molecular and atomic line emission. In particular, energetic electrons excite H$_2$ molecule to vibrational states within the ground electronic state and to electronically excited states. Both excitation paths eventually lead to the infrared emission of H$_2$ in interstellar clouds. Recently, \cite{Bialy2026} reported the detection of H$_2$ ro-vibrational emission excited by cosmic-ray electrons in the starless core Barnard 68 using observations from the James Webb Space Telescope.

The first detailed calculations of fast electrons slowing down in molecular hydrogen gas that took into account ionisation, ro-vibrational and electronic excitation of H$_2$ molecule were carried out by \cite{Jones1973,Olivero1973,Glassgold1973,Cravens1975}. More elaborate and up-to-date analysis was carried out in more recent papers by \cite{Voit1991,Gredel1995,Dalgarno1999}. In all these studies, the sparse experimental data on cross sections were used in the modelling of H$_2$ molecule excitation \citep{Ehrhardt1968,Miles1972,England1988,Buckman1990}. \cite{Shemansky1985} suggested an analytic formula with a number of parameters to approximate the cross sections for excitation to excited electronic states of H$_2$ molecule. In their method, the relative values of the parameters were determined by fitting the experimentally measured relative excitation function. An absolute value of the cross sections was fixed by identifying the Born component in the excitation function. The fitting parameters were assumed to be independent of rotational and vibrational quantum numbers. The lack of accurate cross sections for electron-impact excitation of H$_2$ molecule made it difficult to accurately reproduce experimentally measured energy deposition parameters. \cite{Dalgarno1999} introduced a pseudo-state in their simulations: the magnitude of the excitation cross section of this state was chosen so that there was an agreement between calculated and experimentally measured deposited energy per ion pair.

Recently, the complete set of cross sections for the electron-impact excitation to ro-vibrational levels of the ground and electronically excited states of H$_2$ molecule was calculated using the adiabatic-nuclei molecular convergent close-coupling (MCCC) method \citep{Zammit2017a,Zammit2017b,Scarlett2021}. \cite{Horton2021} conducted Monte Carlo simulations of electron energy deposition in H$_2$ gas using all cross sections generated by MCCC method (for electron-impact excitation, ionisation, and elastic scattering). The mean energy per ion pair obtained in their work was in good agreement with the experimentally determined value. Initially, the cross sections of electron-impact excitation of H$_2$ molecule calculated by \cite{Scarlett2021} were not rotationally resolved, i.e., the rotational structure of the initial and final states was not taken into account. \cite{Scarlett2023} calculated rotationally resolved cross sections for transitions within the vibrational state $v = 0$ of the ground electronic state. Later, these calculations were extended to include all transitions within the ground electronic state, as well as transitions connecting ground and excited electronic states. \cite{Padovani2022} utilised some of these cross sections in the calculations of intensity of H$_2$ ro-vibrational transitions excited by cosmic-ray electrons in interstellar molecular clouds. \cite{Padovani2024} calculated the ultraviolet spectrum resulting from H$_2$ excitation by cosmic-ray electrons, and determined the cosmic-ray induced photodissociation and photoionisation rates of chemical species in molecular clouds. 

Here we consider slowing down of fast electrons in a partially ionised H$_2$ gas. The time-evolution of electron energy distribution is calculated by solving the differential equation. We calculated the yields for electronic excitations, vibrational excitations, dissociation of H$_2$ molecule, and gas heating efficiencies. We compare our results with those of similar studies in which the Monte Carlo approach was used.

\section{Description of the numerical model}\label{sec3}
\subsection{Physical processes and cross sections}
\subsubsection{Electron-impact ionisation of He atoms and H$_2$ molecules}

The approximation formulas for differential cross sections of electron-impact ionisation of atoms and molecules are provided by \cite{Kim1994,Kim2000}. Here, we use relativistic expressions for differential cross sections from the work by \cite{Kim2000}: relativistic effects double the total ionisation cross section of H$_2$ and He at electron energy 300~keV and dominate the cross section at higher electron energies. \cite{Kim1994} provided an analytic approximation of the dipole oscillator strength for H$_2$, He, H, which is used in the cross section formulas. For one-electron ions, \cite{Kim1994} suggested using the same coefficients in the dipole oscillator strength formula as for hydrogen atom. We take into account that a small fraction of H$_2$ ionisations is accompanied by dissociation. We use partial cross sections for the H$_2$ ionisation from the work by \cite{Straub1996}. A set of dissociative ionisation cross sections was also calculated by \cite{Wunderlich2021}. A contribution of dissociative ionisation to the total ionisation cross section is about 7 per cent at the total cross section maximum at electron energy 70~eV. We omit double ionisation of species: the ratio of cross sections of double ionisation and single ionisation is about $0.3-0.5$ per cent for H$_2$ \citep{Kossmann1990}, and about 0.5 per cent for He \citep{Genevriez2019}. 

At low electron energies close to the ionisation threshold of H$_2$ molecule, the calculations are sensitive to the near threshold behaviour of the cross sections. The cross section approximations may have low accuracy in this energy region. We consider the initial electron energies $E \geq 30$~eV.

\subsubsection{Spectroscopic data on H$_2$ molecule} 
\begin{figure}[h]
\centering
\includegraphics[width = 1.\textwidth]{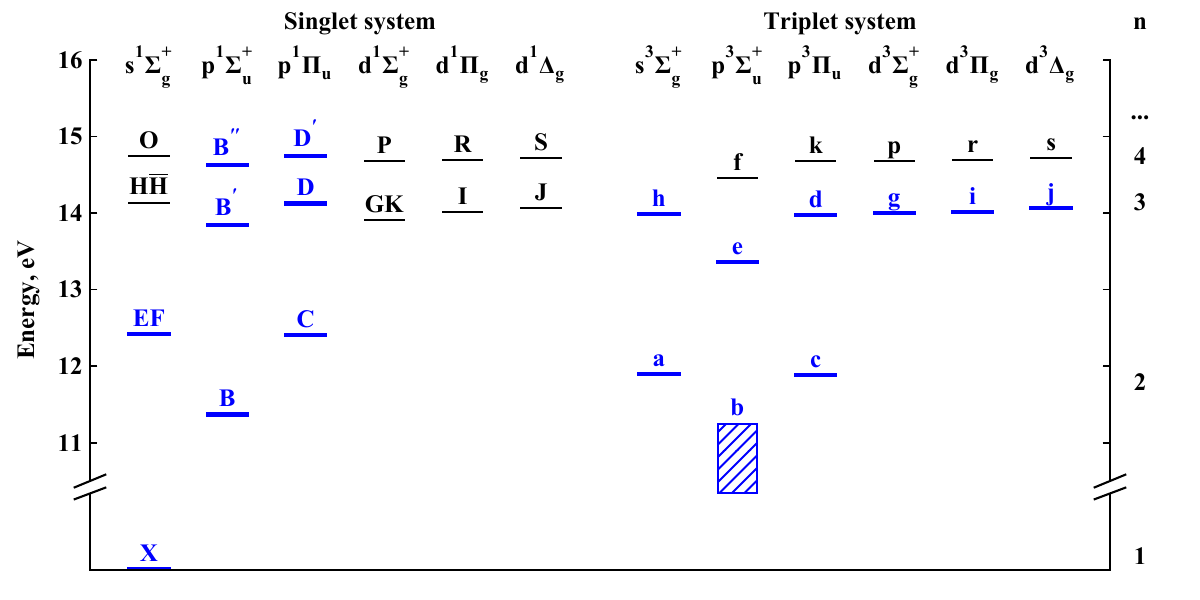}
\caption{H$_2$ electronic states. The states that are taken into account in the simulations are designated by bold blue lines. The energies of electronic states are taken from \cite{Fantz2006}. The triplet state $b$ is a repulsive state.}\label{fig_1}
\end{figure}

Our model takes into account the ro-vibrational energy levels of the following H$_2$ electronic states: the ground electronic state $1s\sigma$ ${}^{1}\Sigma^{+}_g$ (labelled as $X$); the excited singlet electronic states $2p\sigma$ ${}^{1}\Sigma^{+}_{u}$ ($B$), $2p\pi$ ${}^{1}\Pi_{u}$ ($C$), $2s\sigma$ ${}^{1}\Sigma^{+}_{g}$ ($EF$), $3p\sigma$ ${}^{1}\Sigma^{+}_{u}$ ($B^{\prime}$), $3p\pi$ ${}^{1}\Pi_{u}$ ($D$), $4p\sigma$ ${}^{1}\Sigma^{+}_{u}$ ($B^{\prime\prime}$), $4p\pi$ ${}^{1}\Pi_{u}$ ($D^{\prime}$); the triplet states $2s\sigma$ ${}^{3}\Sigma_{g}^{+}$ ($a$), $2p\pi$ ${}^{3}\Pi_{u}$ ($c$), $3p\pi$ ${}^{3}\Pi_{u}$ ($d$), and the repulsive triplet state $2p\sigma$ ${}^{3}\Sigma_{u}^{+}$ ($b$). We also consider excitation to triplet electronic states $3s\sigma$ ${}^{3}\Sigma_{g}^{+}$ ($h$), $3p\sigma$ ${}^{3}\Sigma_{u}^{+}$ ($e$), $3d\sigma$ ${}^{3}\Sigma_{g}^{+}$ ($g$),  $3d\pi$ ${}^{3}\Pi_{g}$ ($i$), $3d\delta$ ${}^{3}\Delta_{g}$ ($j$). The ro-vibrational energy levels of the triplet electronic states $h$, $e$, $g$, $i$, $j$ are not considered, we treat these states as dissociative. Figure~\ref{fig_1} shows the electronic states of H$_2$ molecule up to the principle quantum number $n = 4$ and up to the $d$-shell \citep{Fantz2006}. The states that are taken into account in simulations are designated by bold blue lines in figure~\ref{fig_1}.

We take into account 302 ro-vibrational levels of the ground electronic state $X\,{}^{1}\Sigma^{+}_g$. The energy levels and Einstein coefficients for transitions within the ground electronic state are taken from \cite{Roueff2019}. We take into account 882 ro-vibrational levels of the excited electronic state $B$, 248 levels of $C^{+}$ and 251 levels of $C^{-}$, 108 levels of the state $B^{\prime}$, 28 levels of $D^{+}$ and 336 levels of $D^{-}$. The energies of ro-vibrational levels of excited electronic states $B$, $C$, $B^{\prime}$, $D$, the Einstein coefficients for transitions connecting excited electronic states and the ground state are taken from data files of the CLOUDY code \citep{Abgrall2000,Ferland2017}. The excited electronic states can decay into the vibrational continuum of the ground electronic state. The probabilities of spontaneous radiative dissociation for these states, and the kinetic energy of the produced H atoms, are also taken from the CLOUDY code files \citep{Abgrall2000,Ferland2017}.

The 793 ro-vibrational levels of the electronic state $EF$ with angular momentum quantum number $J \leq 26$ are taken into account. The band transition moments between the excited singlet states $EF$ and $B$ are taken from \cite{Glass-Maujean1984}. These band transition moments take into account the dependence on the vibrational quantum number but neglect dependence on the angular momentum quantum number. We estimate the spontaneous emission probabilities for $R$ and $P$ transitions using the H\"{o}nl--London factors. In fact, the ratio of spontaneous emission probabilities for $R$ and $P$ transitions strongly depends on the mixing between parallel and transverse contributions to the electronic dipole moment function (M. Glass-Maujean, 2025, personal communication). 

We consider 276 ro-vibrational levels of $B^{\prime\prime}$, 72 levels of $D^{\prime+}$ and 182 levels of $D^{\prime-}$. In the case of $B^{\prime\prime}$ and $D^{\prime+}$ states, only energy levels with angular momentum quantum number $J \leq 4$ are taken into account. The radiative transition probabilities for these states were measured and calculated by \cite{Glass-Maujean2007,Glass-Maujean2007b,Glass-Maujean2008,Glass-Maujean2009}. For electronic states $B^{\prime\prime}$ and $D^{\prime}$, decay into the vibrational continuum of the ground electronic state is not taken into account.

We take into account 384 levels of the triplet electronic state $a$, 419 levels of the state $c$, and 412 levels of the state $d$. The energy levels of the triplet electronic state $a$ are taken from \cite{Wolniewicz2007}. In the absence of collisional de-activation, all excitations to triplet states eventually lead to molecule dissociation: the states $c$ and $d$ radiatively decay into the $a$ state, and the $a$ state radiatively decays into the repulsive $b$ state \citep{Avakyan1998,Fantz2006}. For rotational levels of the $c$ state that lie below the $a$ state, the dissociation occurs by coupling to the $b$ state or by dissociative transitions to the $b$ state \citep{Liu2010}. In our calculations of the heat input from H$_2$ dissociation, we assume that the kinetic energy of released H atoms is 3~eV if dissociation occurs by excitation to the bound energy levels of triplet states \citep{Liu2010}. The most probable value of the kinetic energy of H atoms released in dissociative excitation to the repulsive state $b$ is equal to 5.5~eV for the vibrational state $v=0$ of H$_2$ molecule \citep{Trevisan2002}.

\subsubsection{Data on e-H$_2$ scattering -- the molecular convergent close-coupling method}
For the electronic excitation of H$_2$ molecules in collisions with electrons, we use a set of rotationally resolved cross sections calculated using the MCCC method \citep{Zammit2017a,Scarlett2023}. The MCCC method is a fully quantum-mechanical computational method for calculating highly accurate cross sections for electrons, positrons, and (more recently) protons scattering on simple molecules \citep{Plowman2026}. The particular strength of the method is in its computationally-efficient implementation of the close-coupling expansion, allowing for convergence in the cross sections to be explicitly demonstrated via a series of successively larger calculations. With all important reaction channels coupled, the MCCC method can produce self-consistent sets of collision data for many thousands of reactions over a broad range of incident energies from the threshold to the high-energy regime where the Born approximation is valid.

Details of the application of the MCCC method to the calculation of rovibrationally-resolved cross sections for electron-impact excitation of H$_2$ molecule are given in \cite{Scarlett2023}, where results for pure rotational excitation in the ground electronic state were presented. These calculations have since
been extended to include all ro-vibrational transitions within the ground electronic state, and ro-vibrationally-resolved excitation to the singlet electronic states $B$, $C$, $EF$, $B^{\prime}$, $D$ and the triplet electronic states $a$, $c$, and $d$ of H$_2$ molecule. Some of these results have been applied previously by \cite{Padovani2024} in studies of ultraviolet luminescence in molecular clouds. The cross sections for electron-impact excitation to the states $B^{\prime\prime}$ and $D^\prime$ are not yet available. We use scaling factors to calculate these cross sections -- we assume that the cross sections for the excitation to the energy levels of the $B^{\prime\prime}$ state are 0.35 times the $B^{\prime}$ cross sections, and for the excitation to the $D^{\prime}$ energy levels are 0.4 times the $D$ cross sections \citep{Zammit2017b,Padovani2024}. The ro-vibrational transitions connecting the ground electronic state $X$ and excited ungerade $\Sigma$ electronic states have $\Delta J = \pm 1, \pm 3, \pm 5$; transitions connecting the ground state $X$ and excited gerade $\Sigma$ electronic states have $\Delta J = 0, \pm 2, \pm 4$. The transitions between the ground electronic state $X$ and ungerade $\Pi$ states have $\Delta J = 0, \pm 1, \pm 2, \pm 3, \pm 4, \pm 5$ ($\Delta J$ is odd for energy levels with the same total parity and is even otherwise). We take into account dissociative excitation to singlet electronic states and dissociative excitation to the repulsive triplet electronic state $b$ \citep{Scarlett2021}. For triplet electronic states $h$, $e$, $g$, $i$, $j$, we use the total excitation cross sections calculated by \cite{Scarlett2021}. The cross sections are available at electron energies up to 500~eV for the excitation to the singlet electronic states and up to 150~eV for the excitation to the triplet electronic states. We extrapolate excitation cross sections for singlet electronic states at high energies using the power law function, where the exponent is taken equal to that of the ionisation cross section in non-relativistic limit. The excitation cross sections for triplet electronic states decline very fast as the energy increases. For ro-vibrational transitions within the ground electronic state of H$_2$, the cross sections are available for $\Delta J = 0, \pm 2$ and electron energies up to 12~eV.

\subsubsection{Collisions of H$_2$ molecules with neutral species}
We take into account collisions of H$_2$ with H$_2$ and He. For H$_2$--H$_2$ collisions, the data by \cite{Wan2018} are used. The rate coefficients for pure rotational transitions involving rotational levels with $J \leq 31$ of the ground vibrational state of H$_2$ molecule are provided in their work. For transitions involving other H$_2$ energy levels of the ground electronic state, the data by \cite{FlowerRoueff1998,FlowerRoueff1999} are used. For H$_2$--He collisions, the data by \cite{FlowerRoueffZeippen1998} are used.

\subsubsection{Spectroscopic and collisional data on He atoms} 
Our model takes into account 31 energy levels of He atom (singlet and triplet states with principal quantum number $n \leq 4$, we take into account fine-structure components of energy levels separately). The energy levels and radiative transitions of He atom are taken from the NIST database \citep{NIST}. The two-photon decay of the energy level $1s2s$ $^{1}S$ to the ground energy level $1s^2$ $^{1}S$ is taken into account. The Einstein coefficient for this transition is $A = 51$~s$^{-1}$ \citep{Draine2011}. We use cross sections of electron-impact excitation of He atoms evaluated by \cite{Ralchenko2008}. 

We do not consider the electron-impact excitation of H atoms as we focus on the ionisation and excitation of H$_2$ molecules. The abundance of H atoms is small in dense molecular clouds, e.g. \cite{Goldsmith2005}. We do not take into account the ionisation of species by photons emitted by excited states of helium.

\subsubsection{Elastic scattering and Coulomb energy losses}
For electron momentum transfer in elastic collisions with H$_2$ molecules, we use cross sections published by \cite{Yoon2008}. These cross sections are for electron energies up to 100~eV. The cross sections at higher energies are taken from \cite{Dalgarno1999}. For electron momentum transfer with He, we use the experimental data on cross sections by \cite{Crompton1970,Milloy1977}. The cross sections at electron energies $\varepsilon > 12$~eV are taken from \cite{Dalgarno1999}. We take into account Coulomb energy loss: long-range small-angle scattering of high-energy electrons on thermal electrons. We use the approximate formula from \cite{Swartz1971} for the energy decrease per unit time due to Coulomb energy loss. The parameters used in their formula are the thermal electron temperature $T_{e}$ and the number density of thermal electrons $n_e$. In the calculations we set $T_{e} = 100$~K, the corresponding electron energy is $k T_{e} = 8.62 \times 10^{-3}$~eV that is lower than all the excitation energy thresholds. The results are insensitive to the choice of $T_{e}$ for electron energies $\varepsilon >> k T_{e}$.

\subsection{The energy degradation of electrons in the gas}
High-energy electrons ionise atoms and molecules and produce secondary electrons having sufficient energy to produce further ionisation. The time-evolution of the initial energy distribution of electrons is governed by energy loss processes and by the production of secondary electrons during the ionisation. 

\subsubsection{Ionisation}
Consider electron-impact ionisation of a molecule or atom with ionisation potential $I$. Let $E$ be the energy of the projectile electron, and let $\varepsilon$ and $\varepsilon^{\prime}$ be the energies of the ejected and scattered electrons. The electron cannot be identified as the secondary or incident. We arbitrarily call the slower electron of the two the ejected electron and the faster one the scattered electron \citep{Kim1994}. The rate at which the ionisation collisions with target species $i$ contribute to the number density of electrons with energy $\varepsilon$ and $\varepsilon^{\prime}$ is:

\begin{equation}
P_{\rm ion}(E, \varepsilon, \varepsilon^{\prime}) dE d\varepsilon^{\prime} = n_{i} v_{e}(E) \frac{d n_{e}(E)}{dE}  \frac{d \sigma_{i}(E, \varepsilon^{\prime})} {d\varepsilon^{\prime}} dE d\varepsilon^{\prime},
\end{equation}

\noindent
where $n_{i}$ is the number density of target species $i$, $v_{e}(E)$ is the speed of electron with energy $E$, $\varepsilon = E - I - \varepsilon^{\prime}$, and here we assume that $\varepsilon < \varepsilon^{\prime}$. The increase in the number density of electrons in the energy interval $[\varepsilon, \varepsilon + d\varepsilon]$ due to ionisation is equal to the difference between the rate of injection into and the rate of ejection from the energy interval:

\begin{equation}
P_{\rm ion} (\varepsilon) d\varepsilon = \sum_{i} n_{i} \int\limits_{0}^{\infty} d\varepsilon^{\prime} v_{e}(E) \frac{d\sigma_{i}(E, \varepsilon^{\prime})}{d\varepsilon^{\prime}} \frac{dn_{e}(E)}{dE}  d\varepsilon - \sum_{i} n_{i} v_{e}(\varepsilon) \sigma_{i, \rm tot} (\varepsilon) \frac{dn_{e}(\varepsilon)}{d\varepsilon} d\varepsilon,
\end{equation} 

\noindent
where $E = I + \varepsilon + \varepsilon^{\prime}$ inside the integral and $\sigma_{i, \rm tot} (\varepsilon)$ is the total ionisation cross section of specimen $i$ by electron with energy $\varepsilon$.

\subsubsection{Excitation}
Consider electron-impact excitation of H$_2$ molecule, and let $m \to u$ be the transition from the energy level $m$ of the ground electronic state to one of the levels $u$ of the excited electronic state. The target molecule will recoil, gaining a small amount of kinetic energy from the incident electron. This recoil energy is negligible compared to the excitation energy of the transition $\Delta\varepsilon_{um}$ \citep{Horton2021}. The incident electron having energy $E$ before collision possesses energy equal to $\varepsilon = E - \Delta\varepsilon_{um}$ after the collision. The increase in the number density of electrons having the energy in the interval $[\varepsilon, \varepsilon + d\varepsilon]$ is equal to:

\begin{equation}
\begin{split}
P_{\rm H_2, el} (\varepsilon) d\varepsilon & = \sum_{m, u} n_{{\rm H_2}, m} \sigma_{um}(E) v_{e}(E) \frac{d n_{e}(E)}{dE} d\varepsilon  - \\[5pt]
& -\left( \sum_{m, u} n_{{\rm H_2}, m} \sigma_{um}(\varepsilon) \right) v_{e}(\varepsilon) \frac{d n_{e}(\varepsilon)}{d\varepsilon} d\varepsilon,
\end{split}
\end{equation}

\noindent
where $E = \varepsilon + \Delta\varepsilon_{um}$ inside the sum in the first term. In the calculations of ro-vibrational excitation of energy levels of H$_2$ within the ground electronic state, one should take into account electron-impact de-excitation of H$_2$ that leads to the energy gain by electrons (superelastic collisions). In this case, $\Delta\varepsilon_{um} < 0$. The dissociative excitation to the electronic states of H$_2$ molecules and the excitation of helium atoms are treated analogously. The electron-impact (de)-excitation of excited energy levels of He atoms is not considered as the populations of these energy levels are negligibly small.

\subsubsection{Continuous energy loss processes}
The electron elastic scattering on neutral species and Coulomb scattering are treated as continuous energy degradation processes \citep{Voit1991,Xu1991}. The time derivative of the number density of electrons due to momentum transfer in elastic collisions with particles (H$_2$ and He) and due to Coulomb energy loss is:

\begin{equation}
\frac{d}{dt} \frac{dn_{e}(\varepsilon)}{d\varepsilon} = \frac{d}{d \varepsilon} \left( \frac{d n_{e}(\varepsilon)}{d\varepsilon} \left\lvert \frac{dE_{\rm mt}}{dt} \right\rvert + \frac{d n_{e}(\varepsilon)}{d\varepsilon} \left\lvert \frac{dE_{\rm c}}{dt} \right\rvert\right),
\end{equation}

\noindent
where $dE_{\rm c}/dt$ is the energy decrease per unit time of the high-energy electron due to Coulomb energy loss \citep{Swartz1971}, $dE_{\rm mt}/dt$ is the energy decrease per unit time due to elastic collisions:

\begin{equation}
\left\vert \frac{dE_{\rm mt}}{dt} \right\vert = v_{e}(\varepsilon) \varepsilon \sum_{i} \frac{2m_e}{m_{i}} \sigma_{\rm i, mt}(\varepsilon) n_{i},
\end{equation}

\noindent
where $n_i$ and $m_i$ are the number density and mass of gas species $i$, respectively, $\sigma_{\rm i, mt}(\varepsilon)$ is the momentum transfer cross section, $i$ stands for He and H$_2$. 

Figure~\ref{fig_app_1} in appendix~\ref{secA1} shows the energy loss function of electrons in pure H$_2$ gas.

\subsubsection{The differential equation for the energy distribution of electrons}
We assume that electrons are distributed uniformly in the gas cloud, neglecting transport effects. The balance equation for the energy distribution of electrons is:

\begin{equation}
\begin{split}
\frac{d}{dt} \frac{dn_{e}}{d\varepsilon} & = P_{\rm ion}(\varepsilon) + P_{\rm H_2, el}(\varepsilon) + P_{\rm H_2, rovibr}(\varepsilon) + P_{\rm H_2, diss}(\varepsilon) + P_{\rm He}(\varepsilon) + \\[5pt]
& + \frac{d}{d \varepsilon} \left(\frac{dn_{e}(\varepsilon)}{d\varepsilon} \left\vert \frac{dE_{\rm mt}}{dt} \right\vert + \frac{dn_{e}(\varepsilon)}{d\varepsilon} \left\vert \frac{dE_{\rm c}}{dt} \right\vert \right).
\end{split}
\label{eqspencerfano}
\end{equation}

\noindent
For a steady state energy distribution of electrons, the time derivative of the energy distribution is equal to zero. In this case and with the source term, the equation (\ref{eqspencerfano}) is often called Spencer--Fano equation \citep{Spencer1954}.

\subsection{Calculation of level populations of H$_2$ molecules}
In our model, the excitation mechanism of H$_2$ molecules is collisions of high-energy electrons with H$_2$. The energy levels of H$_2$ molecules are depopulated by spontaneous emission and, at high gas density, by collisions with ambient particles (H$_2$ molecules and He atoms). The system of equations that governs the excitation and de-excitation of ro-vibrational energy levels of the ground electronic state of H$_2$ is:

\begin{equation}
\begin{split}
& \frac{dn_l}{dt} = \sum_{m \neq l} \left(\gamma_{lm} + \gamma_{lm}^{\rm eff} \right) n_m - n_{l} \left[ \sum_{m \neq l} \left( \gamma_{ml} + \gamma_{ml}^{\rm eff} \right) + \gamma_{{\rm diss}, l} \right] \\
& + \sum_{m > l} A_{lm} n_{m} - n_{l} \sum_{m < l} A_{ml} + \sum_{m} C_{lm} n_{m} - n_{l} \sum_{m} C_{ml} \\
& - n_l k_{{\rm ion},l} - n_l k_{{\rm diss},l},
\end{split}
\label{eq_h2_stat_equil}
\end{equation}

\noindent
where $n_l$ is the population density of the energy level $l$, $A_{lm}$ is the Einstein coefficient for the spontaneous emission in the transition $m \to l$ within the ground electronic state, $C_{lm}$ is the rate coefficient for the transition from level $m$ to level $l$ through collisions with H$_2$ and He, $k_{{\rm ion}, l}$ is the rate of H$_2$ destruction due to ionisation, $k_{{\rm diss}, l}$ is the rate of H$_2$ destruction due to dissociative excitation or due to excitation to the triplet states of H$_2$ in collisions with electrons, $\gamma_{lm}$ is the rate coefficient for direct ro-vibrational transitions within the ground electronic state induced by collisions with electrons. The electron-impact excitation to the electronic states and subsequent decay of these states are described by the effective transition rates in equation (\ref{eq_h2_stat_equil}): $\gamma_{lm}^{\rm eff}$ is the effective rate of the transition $m \to l$ via excitation to the electronic states, and $\gamma_{{\rm diss}, l}$ is the dissociation rate of H$_2$ molecules in this process.

The rate of direct ro-vibrational excitation from energy level $m$ to level $l$ of the ground electronic state of H$_2$ molecule is:

\begin{equation}
\gamma_{lm} = \int\limits_{\Delta\varepsilon_{lm}}^{\infty} d \varepsilon \, \frac{d n_e(\varepsilon)}{d\varepsilon} v_e(\varepsilon) \sigma_{lm}(\varepsilon),
\label{eq_el_exc_rate}
\end{equation}

\noindent
where $\sigma_{lm}(\varepsilon)$ is the excitation cross section. The effective rate of transition from level $m$ to level $l$ within the ground electronic state of H$_2$ through electron-impact excitation to the excited electronic states is:

\begin{equation}
\displaystyle
\gamma_{lm}^{\rm eff} = \sum_{u} \frac{A_{lu}}{A_{{\rm tot},u}} \gamma_{um}^{\rm el},
\end{equation}

\noindent
where $\gamma_{um}^{\rm el}$ is given by equation (\ref{eq_el_exc_rate}), but in this case $u$ is the energy level of the excited electronic state, $A_{lu}$ is the probability of the transition from the energy level $u$ to $l$ due to spontaneous emission, and $A_{{\rm tot},u}$ is the total decay probability of the energy level $u$ due to spontaneous emission. The total decay probability includes the dissociation probability as a result of the decay into the vibrational continuum of the ground electronic state $A_{{\rm vc}, u}$:

\begin{equation}
\displaystyle
A_{{\rm tot}, u} = \sum_{l} A_{lu} + A_{{\rm vc}, u}.
\end{equation}

\noindent
The effective rate of H$_2$ dissociation (Solomon process) is:

\begin{equation}
\displaystyle
\gamma_{{\rm diss},m} = \sum_{u} \frac{A_{{\rm vc}, u}}{A_{{\rm tot}, u}} \gamma_{um}^{\rm el}.
\end{equation}

The heating rate of the neutral component of the gas due to collisions of ambient atoms and molecules with ro-vibrationally excited H$_2$ molecules is:

\begin{equation}
q_{\rm H_2} = \sum_{m > l} h\nu_{ml} \left( C_{lm} n_{m} - C_{ml} n_l \right),
\label{neutral_gas_heating}
\end{equation}

\noindent
where $C_{lm}$ and $C_{ml}$ are the rates of downward and upward transitions, respectively, in collisions with H$_2$ and He. The low-lying rotational levels of H$_2$ molecule ($J = 2, 3$) have a long decay lifetime. At the end of the simulations, there is non-negligible abundance of H$_2$ molecules at rotational levels $J \geq 2$ that do not have enough time to decay due to the final simulation time. We account for energy locked in these low-lying rotational levels as heat.

\subsection{Numerical Calculations}
The primary electrons have a mono-energetic energy distribution. Initially, the gas is composed of H$_2$ molecules and He atoms, the abundance of thermal electrons in the gas is small, $x_{e} << 1$. The small abundances of ions H$_2^+$, H, H$^+$, He$^{+}$, He$^{++}$ appear as the electrons ionise and excite gas species. The number density of primary electrons is taken to be equal to an arbitrary small value of 10$^{-3}$~cm$^{-3}$. This ensures that the concentration of produced ions and secondary electrons is negligibly small. We perform simulations for the maximal model time $\tau = 10^9$~s. We consider ionisation fractions $x_e = n_e/n_{\rm H, tot} = 0-10^{-2}$, He abundances $x_{\rm He} = n_{\rm He}/ n_{\rm H, tot} = 0-0.1$ and hydrogen molecule number densities $n_{\rm H_2} = n_{\rm H, tot}/2 = 10^4-10^6$~cm$^{-3}$, where $n_{\rm H, tot}$ is the total concentration of H nuclei. The ortho-to-para-H$_2$ ratio is equal to 3, H$_2$ molecules reside in energy levels $J = 0$ and 1. The temperature of the neutral gas component is taken equal to 15~K. The metal ions and dust particles are not taken into account in the simulations. 

The range of electron energies from 0 to maximal electron energy is divided into intervals. Let $n_{i}$ be the electron number density in the energy bin $i$. In appendix~\ref{secA2} we describe the calculations of the time derivatives of variables $n_i$. We solve the system of differential equations for electron number densities in energy bins $n_i$, the number densities of chemical species $e^{-}$, H, H$^+$, H$_2$, H$_2^{+}$, He, He$^{+}$, He$^{++}$, population densities of energy levels of H$_2$ and He. The differential equation system is solved using the SUNDIALS CVODE v5.7.0 equation solver \citep{Hindmarsh2005,Gardner2022}.

\begin{figure}[h]
\centering
\includegraphics[width = 0.6\textwidth]{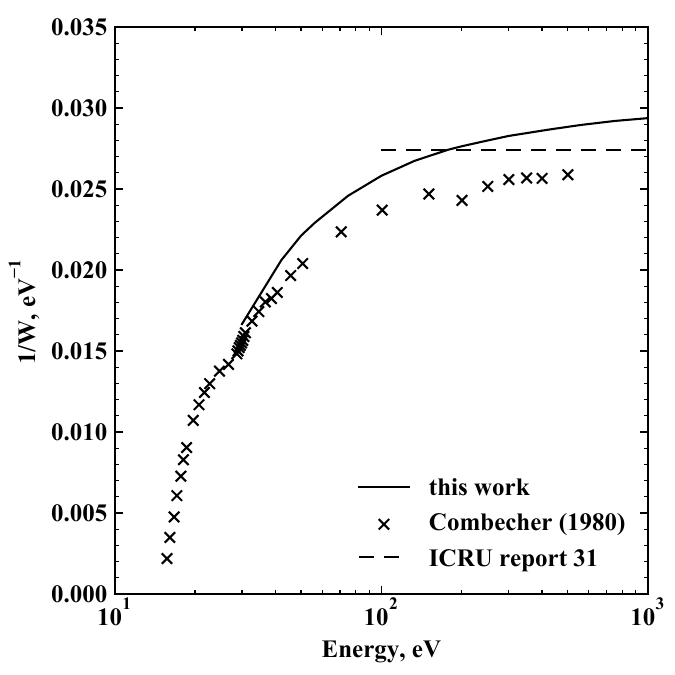}
\caption{Inverse of the energy per ion pair for pure H$_2$ gas. The solid line is the results of our simulations, the crosses represent the measurements by \cite{Combecher1980}, the high energy value recommended by the ICRU is shown by the dashed line \citep{ICRU_31}. The total experimental error stated by  \cite{Combecher1980} is less than 2 per cent.}\label{fig_2}
\end{figure}

\section{Results}\label{sec4}
\subsection{The mean energy per ion pair}
The mean deposited energy per ion pair $W$ is equal to the initial energy of the primary electron divided by the total number of ionisations produced by the primary and secondary electrons. The parameter $W$ approaches a constant asymptotic value as the primary electron energy increases. Figure~\ref{fig_2} shows the results of our simulations of the mean energy per ion pair $W$ for pure neutral H$_2$ gas. The asymptotic value of $W$ recommended by the International Commission on Radiation Units and Measurements (ICRU) is $36.5 \pm 0.3$~eV \citep{ICRU_31}. This recommendation is based on various measurements of the parameter, and is an average over a wide range of primary electron energies, e.g. \cite{Jesse1955,Weiss1956}. We also show the experimentally determined values of the mean energy per ion pair $W$ from \cite{Combecher1980}. Our calculations yield a value of 34.7~eV at electron energy of 0.5~keV. This value is approximately 10 per cent lower than the value of 38.64~eV determined experimentally by \cite{Combecher1980} at the same energy and approximately 5 per cent lower than the parameter value recommended by ICRU.

The difference in the calculations and the experimental data may be explained by: i) the uncertainties in the cross section data set; ii) incomplete set of H$_2$ electronic states used in the simulations; iii) uncertainty in the calculations of electron energy loss by excitation to the triplet state $b$:

i) Uncertainties in the MCCC cross sections for H$_2$ excitation are estimated to be about 10 per cent \citep{Zammit2017b}. \cite{Horton2021} conducted Monte Carlo simulations of electron energy deposition in H$_2$ gas, and studied the propagation of uncertainties in the cross sections through simulations using the fast Total Monte Carlo method \citep{Rochman2014}. The uncertainties in cross sections lead to uncertainties of the calculated mean energy per ion pair of the same order of magnitude. 

ii) The larger the number of H$_2$ electronic states is taken into account, the higher the energy losses by the H$_2$ excitation and the higher the mean energy per ion pair. We have taken into account two electronically excited states $B^{\prime\prime}$ and $D^{\prime}$ with the principal quantum number $n = 4$. The excitation cross sections for these states are taken equal to the corresponding cross sections for states $B^{\prime}$ and $D$ multiplied by scaling factors. The contribution of states $B^{\prime\prime}$ and $D^{\prime}$ to the total electron energy loss is about 1.5 per cent. The contribution of triplet electronic states with the principal quantum number $n = 3$ ($h$, $e$, $d$, $g$, $i$, $j$) to the electron energy loss is approximately 1 per cent.

iii) The energy lost by an electron in excitation to the repulsive triplet state $b$ is equal to the sum of the bond energy of H$_2$ molecule $\Delta\varepsilon = 4.5$~eV and the kinetic energy of H atoms $\varepsilon_{\rm kin}$ released in dissociation. The kinetic energy $\varepsilon_{\rm kin}$ has a wide distribution that depends on the incident electron energy and the vibrational quantum number of the energy level of H$_2$ \citep{Trevisan2002}. The parameter $\varepsilon_{\rm kin}$ has a lower boundary equal to approximately 2.5~eV for the vibrational quantum number $v = 0$: the cross section is negligibly small at $\varepsilon_{\rm kin} < 2.5$~eV \citep{Trevisan2002,Scarlett2021}. We take this lower boundary of the kinetic energy of H atoms in the calculations of electron energy loss. According to our calculations, the contribution of excitation to the triplet state $b$ to the electron energy loss is approximately 5 per cent at high initial electron energies.

\begin{figure}[h]
\centering
\includegraphics[width = 1.\textwidth]{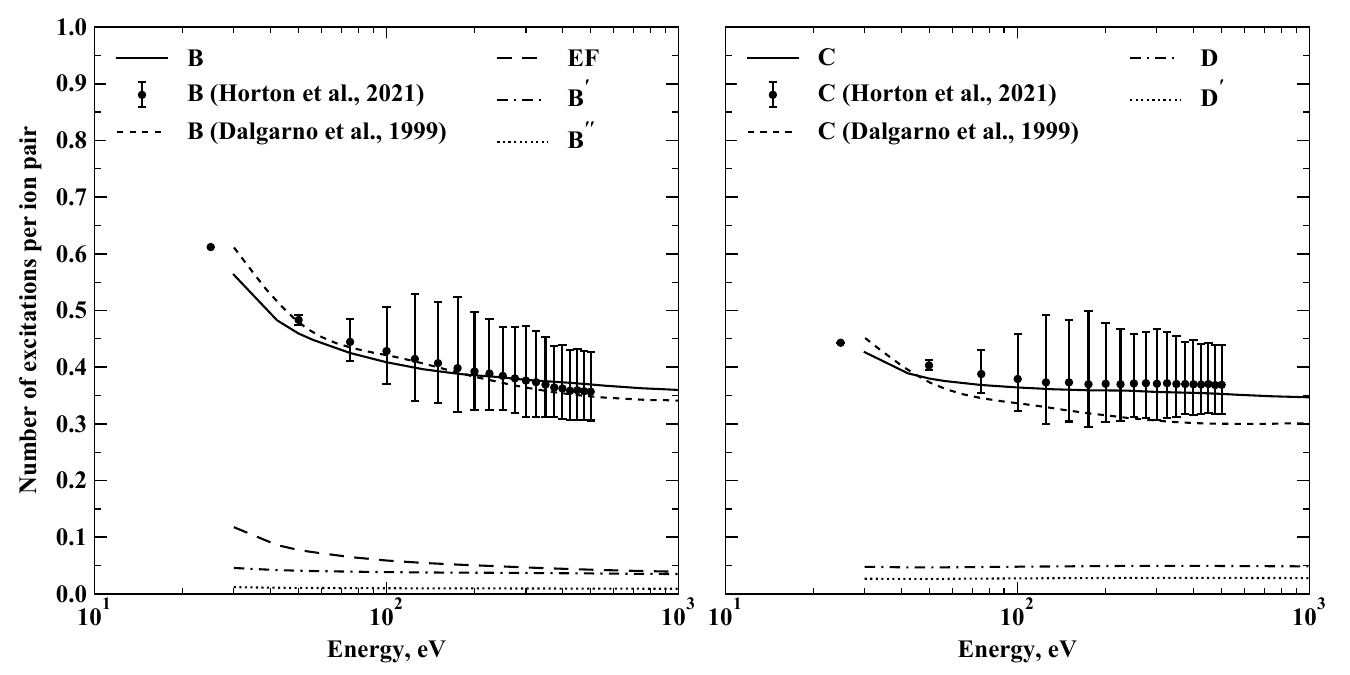}
\caption{The number of excitations per ion pair of H$_2$ electronic states for pure neutral H$_2$ gas. The results of calculations by \cite{Dalgarno1999} and \cite{Horton2021} are presented for electronic states $B\,{}^{1}\Sigma^{+}_{u}$ and $C\,{}^{1}\Pi_{u}$ for comparison. The results by \cite{Horton2021} are shown with the uncertainties estimated using the fast Total Monte Carlo method \citep{Rochman2014}.} 
\label{fig_3}
\end{figure}

\subsection{The electronic excitation of H$_2$ molecules}
Figure~\ref{fig_3} presents the calculated excitation number per ion pair for electronic states of H$_2$ molecule for pure neutral H$_2$ gas. For comparison, the calculations by \cite{Dalgarno1999} and \cite{Horton2021} are also shown. We only include direct excitation to H$_2$ electronic states in the results presented in figure~\ref{fig_3}, similar to \cite{Horton2021}. \cite{Dalgarno1999} took into account the cascading from higher lying singlet electronic states in their results. In our simulations and those by \cite{Horton2021}, the cross sections for H$_2$ electron-impact excitation generated by the MCCC method are used. We solve the differential equation for the electron energy distribution in our modelling, while \cite{Horton2021} used Monte Carlo simulations. There is a good agreement between the results of our simulations and those by \cite{Horton2021}. The MCCC cross sections for excitation to the $C$ state are about 1.5 times larger than the cross sections used by \cite{Dalgarno1999} at electron energies $E > 50$~eV. It explains the difference between the results of our simulations and those by \cite{Dalgarno1999} for the $C$ state, see also \cite{Horton2021}.

The calculated mean excitation number per ion pair is 0.36 for the $B$ state and 0.35 for the state $C$ for the initial electron energy of 1~keV. The excitation number of the electronic state $EF$ is 0.04 per ion pair. The excitation to the electronic state $EF$ is followed by the cascading to the electronic state $B$. Thus, the $EF - B$ cascade contribution to the total internally generated flux of Lyman band photons is about 10 per cent. The excitation numbers per ion pair for other electronic states are 0.035, 0.009, 0.049, and 0.028 for $B^{\prime}$, $B^{\prime\prime}$, $D$ and $D^{\prime}$, respectively, for the initial electron energy of 1~keV. The results can be expressed in terms of the mean energy per excitation that is equal to the ratio of the mean energy per ion pair and the mean excitation number per ion pair. According to our calculations, the mean energy per excitation is about 95 and 98~eV for the $B$ and $C$ states at 1~keV, respectively. For electronic states $B^{\prime\prime}$ and $D^{\prime}$, this parameter is 3.7~keV and 1.2~keV, respectively. 

\begin{figure}[h]
\centering
\includegraphics[width = 0.6\textwidth]{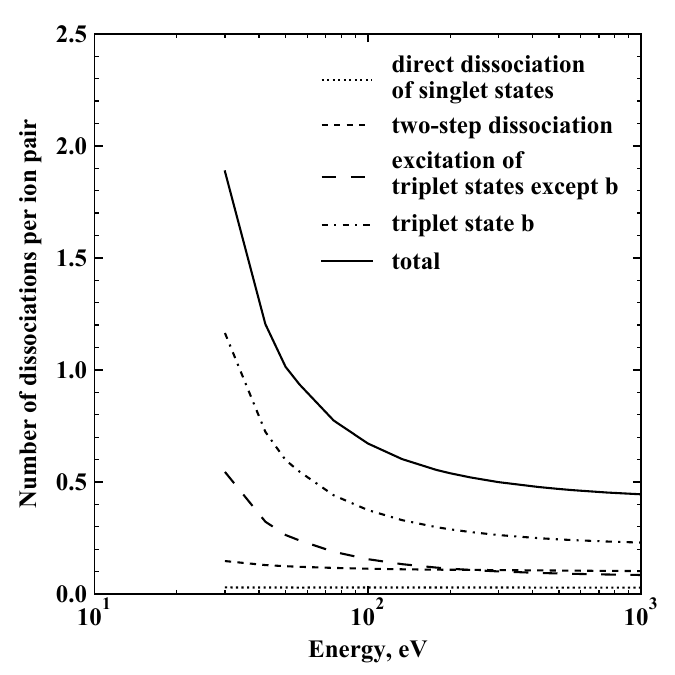}
\caption{The number of H$_2$ dissociations per ion pair for pure neutral H$_2$ gas.} 
\label{fig_4}
\end{figure}

Dissociation of H$_2$ molecule occurs through: i) dissociative excitation to the singlet electronic states; ii) excitation to the bound levels of excited electronic states and subsequent spontaneous radiative dissociation into the vibrational continuum of the ground electronic state (two-step dissociation or Solomon process); iii) dissociative transitions to the repulsive triplet electronic state $b$; iv) excitation to the triplet electronic states that eventually leads to dissociation of H$_2$ molecules. Figure~\ref{fig_4} shows the number of dissociations per ion pair through each of these processes for pure neutral H$_2$ gas. The most important dissociation path is the excitation to the triplet electronic states, which agrees with the conclusion by \cite{Dalgarno1999}. The dissociative excitation to the triplet state $b$ and excitation to the bound energy levels of triplet states are the main electron energy loss process at electron energies from about 9~eV to the ionisation threshold of H$_2$ molecule, see figure~\ref{fig_app_1}. The total number of dissociations is 0.45 per ion pair at initial electron energy of 1~keV. \cite{Dalgarno1999} reported the parameter value of 0.47.

\subsection{The ro-vibrational excitation of H$_2$ molecules}
The electron-impact excitation to vibrational states of the ground electronic state of H$_2$ proceeds in two ways: i) direct ro-vibrational excitation within the ground electronic state; ii) excitation to the electronic states and subsequent radiative de-excitation back to the ground electronic state (cascading). The excitation numbers per ion pair for these two processes are given in Table~\ref{tabel1}. The results are presented for neutral H$_2$--He gas and two initial electron energies 0.1 and 1~keV, He/H$_2$ = 0.2. For each vibrational quantum number $v \leq 4$, the first line is for the excitation path via radiative cascading from excited electronic states, the second line is for the direct ro-vibrational excitation. For higher vibrational states $v > 4$, only excitation numbers by cascading are given. The entry probabilities $b(v)$ into the vibrational states by \cite{Gredel1995} are also given for comparison. The number of excitations via radiative cascading from excited electronic states is of the order of 0.1 per ion pair for $v=1$, and slowly decreases with vibrational quantum number increase. The direct excitation rate to the state $v=1$ is almost two orders of magnitude higher than the excitation rate by cascading. The cross sections for direct ro-vibrational transitions fall off quickly with the increase in the vibrational quantum number of the final state. For energy levels with $v \geq 3$, the population of vibrational states occurs mainly via radiative cascading from excited electronic states. Our results for direct ro-vibrational excitation are similar to those by \cite{Gredel1995}. \cite{Gredel1995} used cross section data for ro-vibrational excitation derived by \cite{Buckman1990,Rescigno1993}. We obtain excitation numbers through cascading about $1.5-2$ times higher than the simulation results by \cite{Gredel1995}. The difference between our results and those by \cite{Gredel1995} is explained by different cross section data used for electronic excitation of H$_2$ molecule. In appendix~\ref{secA3}, we present excitation numbers per ion pair separately for para-H$_2$ and ortho-H$_2$ for primary electron energies ranging from 30~eV to 1~keV.

\begin{table}[h]
\caption{Number of excitations per ion pair}\label{tabel1}
\begin{tabular}{@{}lccc@{}}
\toprule
& \multicolumn{2}{@{}c@{}}{this work} & \cite{Gredel1995} \\
$v$ & $E$ = 0.1~keV  & 1~keV & $E$ = 0.1~keV \\
\midrule
    1 & 0.078 & 0.069 & 0.035 \\ 
      & 5.99  & 4.32  & 5.04  \\ 
    2 & 0.070 & 0.062 & 0.032 \\ 
      & 0.34  & 0.25  & 0.35 \\ 
    3 & 0.072 & 0.063 & 0.031 \\ 
      & 0.031 & 0.022 & - \\ 
    4 & 0.070 & 0.061 & 0.030 \\ 
      & 0.004 & 0.003 & - \\ 
    5 & 0.068 & 0.060 & 0.028 \\ 
    6 & 0.061 & 0.054 & 0.027 \\ 
    7 & 0.051 & 0.046 & 0.025 \\ 
    8 & 0.044 & 0.039 & 0.022 \\ 
    9 & 0.038 & 0.033 & 0.020 \\ 
   10 & 0.033 & 0.029 & 0.019 \\ 
   11 & 0.031 & 0.026 & 0.018 \\ 
   12 & 0.028 & 0.024 & 0.019 \\ 
   13 & 0.024 & 0.021 & 0.015 \\ 
   14 & 0.016 & 0.014 & 0.010 \\ 
\botrule
\end{tabular}
\footnotetext{The entry probabilities $b(vJ)$ by \cite{Gredel1995} are normalized in such a way that $b(vJ) \zeta n_{\rm H_2}$ is the entry rate in cm$^{-3}$~s$^{-1}$ into individual energy level $(v,J)$, where $\zeta$ is ionisation frequency in s$^{-1}$. We summed over final rotational levels, and assume the ortho-to-para ratio equal to 3. Formally, the definition of entry probability coincides with excitation number per ion pair calculated in our work. The first line is for cascading from excited electronic states for all vibrational quantum numbers $v$, the second line for $v \leq 4$ is for direct ro-vibrational excitation.}
\end{table}

\begin{figure}[h]
\centering
\includegraphics[width = 0.8\textwidth]{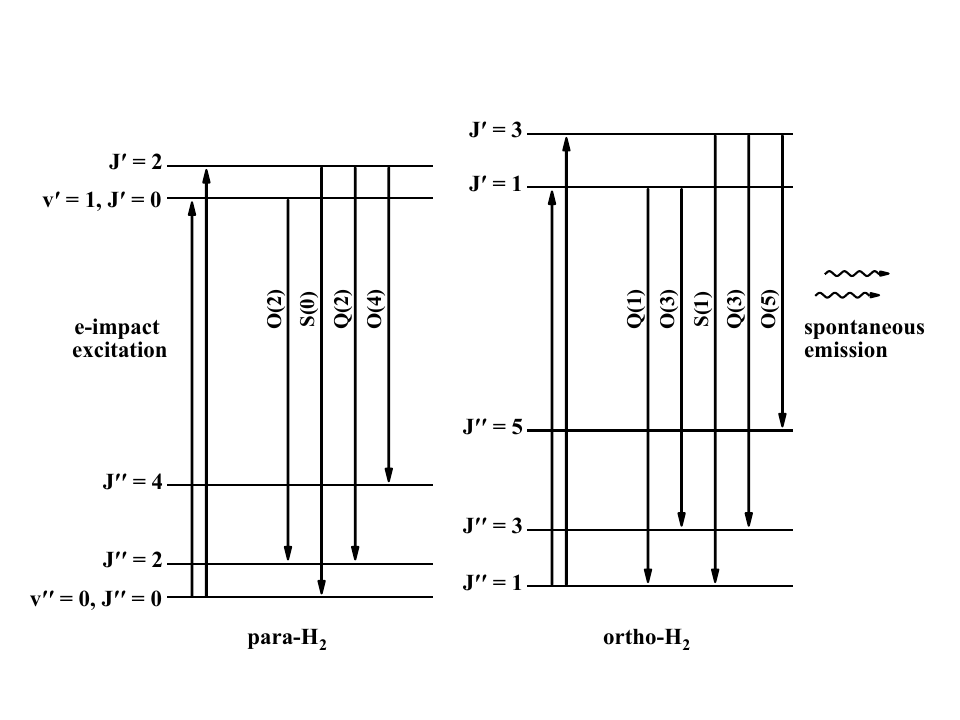}
\caption{The scheme of H$_2$ pumping. Hydrogen molecules are excited to the first vibrational state $v=1$ by electron impact. The subsequent spontaneous decay of excited energy levels leads to the near-infrared photon emission.} 
\label{fig_5}
\end{figure}

The fraction of primary electron energy that is lost in excitation of ro-vibrational transitions of the ground electronic state of H$_2$ molecules (including pure rotational excitation) is 0.11 for a 1~keV electron in pure H$_2$ gas. The fraction of primary electron energy that is lost in pure rotational excitation is 0.034, and on excitation to vibrational states is 0.077. The value of the last parameter agrees with the value calculated by \cite{Dalgarno1999}. The fraction of primary electron energy that is lost in excitation of ro-vibrational transitions $v = 0 \to 1$ is 0.068 for a 1~keV electron. Due to the low temperature of the gas, hydrogen molecules in molecular clouds are preferentially found in the lowest energy levels $(v,J) = (0,0)$ and $(0,1)$. For para-H$_2$, the transitions with $\Delta J = J^{\prime} - J^{\prime\prime} = 0$ and 2 have approximately equal excitation cross sections. The subsequent radiative de-excitation of $v = 1$ energy levels leads to the emission of para-H$_2$ transitions (1–0)O(2), (1–0)Q(2), (1–0)S(0), (1–0)O(4), see figure~\ref{fig_5}. For ortho-H$_2$, the excitation cross section for the transition with $\Delta J = 0$ is about 2 times larger than the cross section for the transition with $\Delta J = 2$. The subsequent radiative de-excitation of $v = 1$ energy levels leads to the emission of ortho-H$_2$ transitions (1–0)Q(1), (1–0)O(3), (1–0)Q(3), (1–0)S(1), (1–0)O(5), see figure~\ref{fig_5}. In dense molecular clouds, the ortho-/para-H$_2$ ratio may be low, and most of H$_2$ molecules are in the para-state, e.g. \cite{Pagani2009}. In this case, the electron energy lost in ro-vibrational excitation of H$_2$ molecules is re-emitted in the para-H$_2$ transitions. The line (1–0)O(2) is approximately $2.5$ times stronger than other $v = 1 \to 0$ para-H$_2$ transitions, see also \cite{Padovani2022,Bialy2026}. The de-excitation of vibrationally excited energy levels of H$_2$ by collisions with ambient gas species is negligible at gas temperatures and densities characteristic of cold molecular clouds \citep{Tine1997,Lique2015}.

The fraction of primary electron energy that is lost in excitation to the singlet electronic states of H$_2$ molecule is about 35 per cent for a 1~keV electron. Most of this energy is emitted as Lyman and Werner band photons. 

\begin{figure}[h]
\centering
\includegraphics[width = 0.6\textwidth]{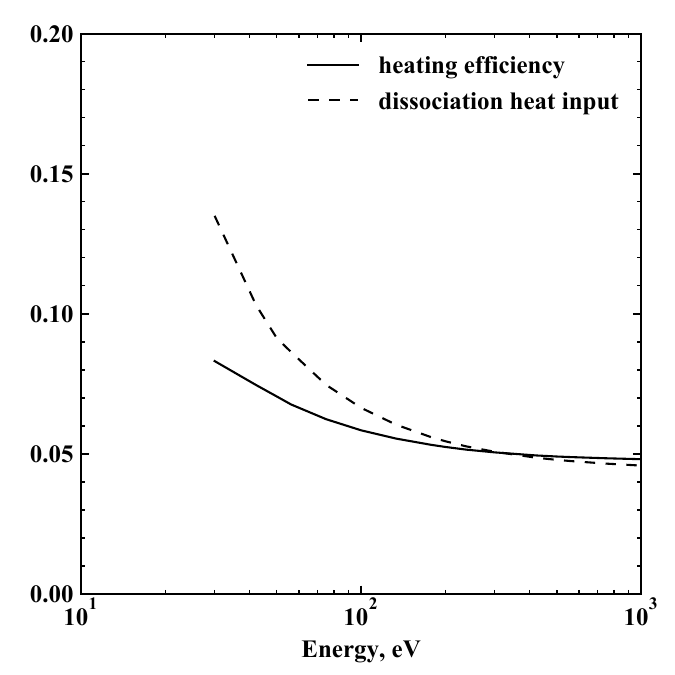}
\caption{Heating efficiency and dissociation heat input as a function of primary electron energy. The results are shown for pure neutral H$_2$ gas with hydrogen molecule number density $n_{\rm H_2} = 10^4$~cm$^{-3}$.} 
\label{fig_6}
\end{figure}

\subsection{The heating of the gas}
The heating efficiency $\eta$ is defined as the fraction of the primary electron energy that is converted into heat from: (i) elastic scattering on H$_2$ and He; (ii) Coulomb energy loss (in the case of partially ionised gas); (iii) collisions of ambient particles with ro-vibrationally excited H$_2$ molecules. The last heating mechanism depends on the gas temperature and density, see eq. (\ref{neutral_gas_heating}). If the gas is warm and dense enough, the low-lying rotational levels of H$_2$ are thermalised, and the electron energy gone to excite these energy levels is converted into heat \citep{Dalgarno1999}. Spontaneous radiative dissociation of excited singlet electronic states, and excitation to the triplet electronic states of H$_2$ molecule also contribute to the heating of the gas. As in \cite{Dalgarno1999}, we take into account this source of heating separately. The dissociation heat input $\xi$ is defined as the fraction of primary electron energy that is converted into heat by this mechanism. The dissociation heat input is dominated by excitation to the triplet electronic states of H$_2$ molecule.

Figure~\ref{fig_6} shows the heating efficiency and dissociation heat input for pure neutral H$_2$ gas as a function of the initial electron energy for the hydrogen molecule number density $n_{\rm H_2} = 10^4$~cm$^{-3}$. The calculated heating efficiency and dissociation heat input are $\eta = 0.048$ and $\xi = 0.046$, respectively, at electron energy of 1~keV. We conducted simulations with hydrogen molecule number density $n_{\rm H_2} = 10^6$~cm$^{-3}$: the heating efficiency is $\eta = 0.056$ and dissociation heat input is $\xi = 0.046$ at electron energy of 1~keV. \cite{Dalgarno1999} calculated heating efficiency and dissociation heat input to be equal to $\eta = 0.057$ and $\xi = 0.053$ for a 1~keV electron, respectively, that are close to the values derived here. \cite{Dalgarno1999} assumed in their modelling that all the energy lost in excitation to rotational levels is entirely converted into heat. 

The fast electrons slowing down in interstellar gas produce ions that participate in exothermic chemical reactions. The chemical heating must be taken into account in the accurate calculation of the gas heating rate \citep{Glassgold2012}.

\subsection{The stopping time of electrons}
\begin{figure}[h]
\centering
\includegraphics[width = 0.6\textwidth]{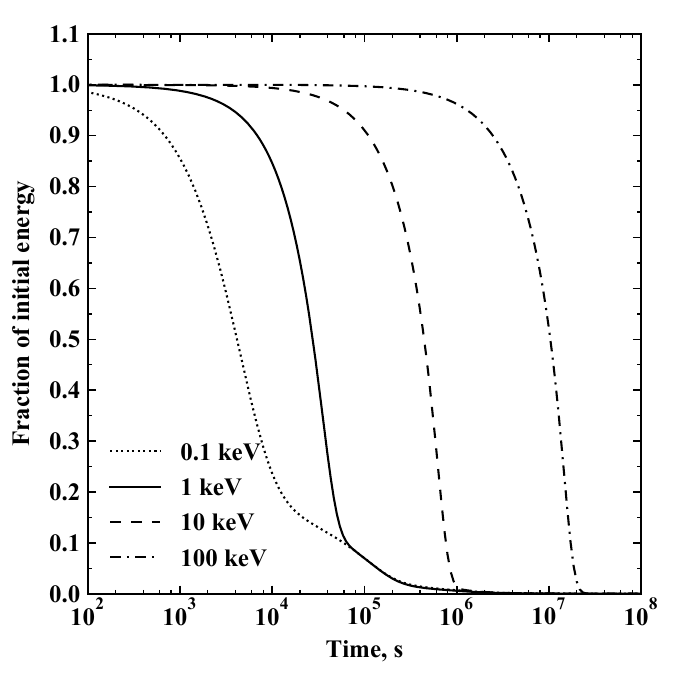}
\caption{The total electron energy (carried by primary and secondary electrons) as a function of time. The fraction of initial energy is along the $y$-axis. The results for initial electron energies 0.1, 1, 10, 100~keV are shown.} 
\label{fig_7}
\end{figure}

We calculated the total electron energy carried by electrons as a function of time. The results are presented in figure~\ref{fig_7} for pure neutral H$_2$ gas. The higher the initial electron energy, the longer the electron stopping time. The speed at which the fast electron loses energy in the medium is characterised by the energy loss function $L_{e}$, see appendix~\ref{secA1}. We can estimate the stopping time of electrons having the energy $\varepsilon$:

\begin{equation}
\tau_{\varepsilon} \approx \frac{\varepsilon}{n_{\rm H_2} L_{e} v_{e}} = 5 \times 10^4 \left(\frac{10^4 \,\text{cm}^{-3}}{n_{\rm H_2}}\right) \left( \frac{\varepsilon}{1~\text{keV}} \right)^{1/2} \left( \frac{ 10^{-15} \,\text{eV cm}^2}{L_{e}} \right) \, \text{s},
\end{equation} 

\noindent
where we assume that $\varepsilon << m_{e} c^2$. In molecular gas with hydrogen number density $n_{\rm H_2} = 10^4$~cm$^{-3}$, the stopping time of electrons with energy $\varepsilon = 1$~keV is about $5\times 10^4$~s. For electrons with energy $\varepsilon = 100$~keV, this time scale is $\tau_{\varepsilon} \approx 2 \times 10^7$~s. It is exactly the result we obtain in numerical simulations, see figure~\ref{fig_7}.

In the case of initial electron energies $E = 0.1$ and 1~keV, there is a change in the slope of the electron energy dependence on time, see figure~\ref{fig_7}. It happens when the average electron energy becomes lower than the ionisation energy threshold of H$_2$ molecule. In this regime, the main process of electron energy losses becomes the excitation to ro-vibrational levels of H$_2$ molecule. The energy loss by ro-vibrational excitation is substantially lower than energy losses by ionisation and excitation to electronic states of H$_2$ molecule, see figure~\ref{fig_app_1}. For high initial electron energies this change of slope also occurs, but it is not seen on figure~\ref{fig_7} for the linear scale of $y$-axis.

\subsection{Energy deposition in partially ionised H$_2$ -- He gas mixture}
We calculated the parameters of electron energy deposition in a gas mixture of H$_2$ and He with various ionisation fractions and He/H$_2$ = 0.2. The results are presented in Table~\ref{table2} and in Tables in appendix~\ref{secA4}. Here, the mean energy per ion pair is equal to the initial energy of electrons $E$ divided by the number of secondary electrons produced. The yields of the excitation and dissociation of H$_2$ molecule are presented in numbers per H$_2^{+}$/H$^{+}$ ions. The presence of helium has little effect on most of the energy deposition parameters, see Table~\ref{table2}.

The electron energy deposition parameters can be divided into three groups having low, moderate and high sensitivity to the ionisation fraction. The Coulomb energy loss is important at low electron energies, see figure~\ref{fig_app_1}. Thus, the parameter sensitivity to the ionisation fraction depends on the energy of electrons generating the physical process for which the parameter is responsible. The cross sections of excitation to singlet electronic states of  H$_2$ have a maximum at $50-100$~eV. The Coulomb energy loss is small at these energies at $x_{e} \leq 10^{-3}$. Thus, the effect of the ionisation fraction on the excitation to singlet electronic states of H$_2$ is negligible at $x_{e} \leq 10^{-3}$. The same is true for other physical processes with high characteristic energies. On the other hand, the fraction of electron energy lost in excitation to ro-vibrational energy levels of H$_2$ molecule, and the heating efficiency have high sensitivity to the ionisation fraction, see Table~\ref{table2}. The yields for excitation to triplet electronic states also become sensitive to the ionisation fraction at $x_{e} \geq 10^{-4}$. It leads to a decrease in the H$_2$ dissociation yield and in dissociation heat input with increasing ionisation fraction -- we designated these parameters as having moderate sensitivity to ionisation fraction.

The observed intensity ratio of lines in the $v = 2 \to 1$ and $v = 1 \to 0$ transitions may be used as a diagnostic of H$_2$ excitation mechanism in molecular clouds \citep{Gredel1995}. The direct ro-vibrational excitation by collisions with electrons determines the populations of low-lying vibrational states. The ratio of electron-impact excitations into $v = 2$ and $v = 1$ vibrational states is 0.057 at high electron energies and low ionisation fraction. This parameter increases slowly with the ionisation fraction. The contribution of direct ro-vibrational transitions to the population of H$_2$ vibrational states $v=1,2$ becomes less important than the electronic excitation followed by radiative cascading at high ionisation fractions, $x_e \geq 10^{-2}$, see appendix~\ref{secA4}. 

The mean energy for He$^{+}$ production is 430~eV for a 1~keV electron at low ionisation fraction. The result by \cite{Dalgarno1999} is 459~eV and the result by \cite{Voit1991} is 441~eV. The fraction of initial electron energy that goes into He excitation is 2 per cent.

\begin{table}[h]
\caption{Parameters of the electron energy degradation in H$_2$ and He gas mixture.}\label{table2}
\begin{tabular*}{\textwidth}{@{}p{5.6cm}llllll@{}}
\toprule
& pure neutral & \multicolumn{4}{@{}c@{}}{H$_2$--He gas mixture} \\
\cmidrule{3-5}
Parameter & H$_2$ gas & $x_e = 0$ & $10^{-7}$ & $10^{-5}$ & $10^{-3}$ \\
\midrule
\multicolumn{6}{@{}c@{}}{\it Low sensitivity to ionisation fraction} \\
Mean energy per ion pair, $W$ [eV]   & 34.0  & 34.7 & 34.7 & 34.8 & 36.1 \\
Number of excitations to $B$ per H$_2^{+}$/H$^{+}$ ion & 0.36 & 0.36 & 0.36 & 0.36 & 0.35 \\
Number of excitations to $C$ per H$_2^{+}$/H$^{+}$ ion & 0.35 & 0.35 & 0.35 & 0.35 & 0.34 \\
Ratio of excitations $v = 2/v = 1$  & 0.057 & 0.057 & 0.058 & 0.068 & 0.077 \\
Mean energy per He$^{+}$ ion [eV] & -- & 430 & 430 & 430 & 438 \\
\multicolumn{6}{@{}c@{}}{\it Moderate sensitivity to ionisation fraction} \\
Number of H$_2$ dissociations per H$_2^{+}$/H$^{+}$ ion & 0.45 & 0.48 & 0.48 & 0.46 & 0.29 \\
Dissociation heat input, $\xi$ & 0.046 & 0.046 & 0.045 & 0.043 & 0.020 \\
\multicolumn{6}{@{}c@{}}{\it High sensitivity to ionisation fraction} \\
Fraction of energy lost in ro-vibrational excitation of H$_2$ & 0.11  & 0.11 & 0.10 & 0.050 & 0.0024 \\
Fraction of energy lost in excitation to the $v=1$ state of H$_2$ & 0.068 & 0.067 & 0.065 & 0.032 & 0.0013 \\
Heating efficiency, $\eta$           & 0.048 & 0.048 & 0.051 & 0.091 & 0.20 \\
\botrule
\end{tabular*}
\footnotetext{The results are shown for initial electron energy 1~keV, hydrogen molecule number density $n_{\rm H_2} = 10^4$~cm$^{-3}$, He/H$_2$ = 0.2.}
\end{table}

\section{Conclusions}\label{sec5}
The rotationally resolved electron-impact excitation cross sections calculated using the MCCC method made possible the accurate modelling of electron energy deposition in interstellar gas. We consider all important energy loss processes of energetic electrons in H$_2$--He gas. Initially, the electron energy distribution is assumed to be mono-energetic, and the time-evolution of the energy distribution is calculated. We solve the non-stationary differential equation of Spencer--Fano type for the electron energy distribution \citep{Spencer1954}. The calculated energy deposition parameters are compared with the results of similar studies in which the Monte Carlo approach was used \citep{Gredel1995,Dalgarno1999,Horton2021}. There is a good agreement with the results by \cite{Horton2021} that used the MCCC cross section data. We have shown that about 11 per cent of the initial energy of fast electrons goes into the ro-vibrational excitation of H$_2$ energy levels (including pure rotational excitation) in neutral molecular gas. The calculated excitation numbers per ion pair for direct excitation to the vibrational states $v = 1$, 2 agree with the results by \cite{Gredel1995}. However, calculated yields for excitation to the vibrational states by radiative cascading from electronically excited states are about $1.5-2$ times higher than the results obtained by \cite{Gredel1995}. The difference between the two simulations is explained by the difference in the cross section data. We found that the downward transitions from the $EF$ electronic state to the $B$ electronic state contribute about 10 per cent to the total number of excitations to the $B$ electronic state.

The approach used in our work has the advantage that it can be incorporated into more sophisticated codes studying the astrophysical phenomena, for example, specimen excitation in interstellar clouds in the vicinity of gamma-ray and X-ray sources, e.g. \cite{Nesterenok2024,Nesterenok2024b}.

\bmhead{Acknowledgements}
The author thanks Liam Scarlett, Mark Zammit, Igor Bray, and Dmitry Fursa for providing cross section data on electron-impact excitation of H$_2$ molecule. The author also thanks Liam Scarlett for help in writing the section on cross section data, and Dmitry Fursa for insightful comments on the draft of the manuscript. The data on energy levels of electronic states $B^{\prime\prime}$, $D^{\prime}$ of H$_2$ molecule, as well as Einstein coefficients for transitions connecting these states and the ground electronic state $X$ were kindly provided by Dr. Mich{\`e}le Glass-Maujean. Nesterenok A.V. was supported by the baseline project FFUG-2024-0002 at the Ioffe Institute.

\section{Data availability statement}
The set of cross sections for electron-impact excitation of H$_2$ molecule can be downloaded from the MCCC database website at http://mccc-db.org .

\begin{appendices}
\section{Energy loss function for electrons}\label{secA1}
\begin{figure*}[h]
\centering
\includegraphics[width = 1.\textwidth]{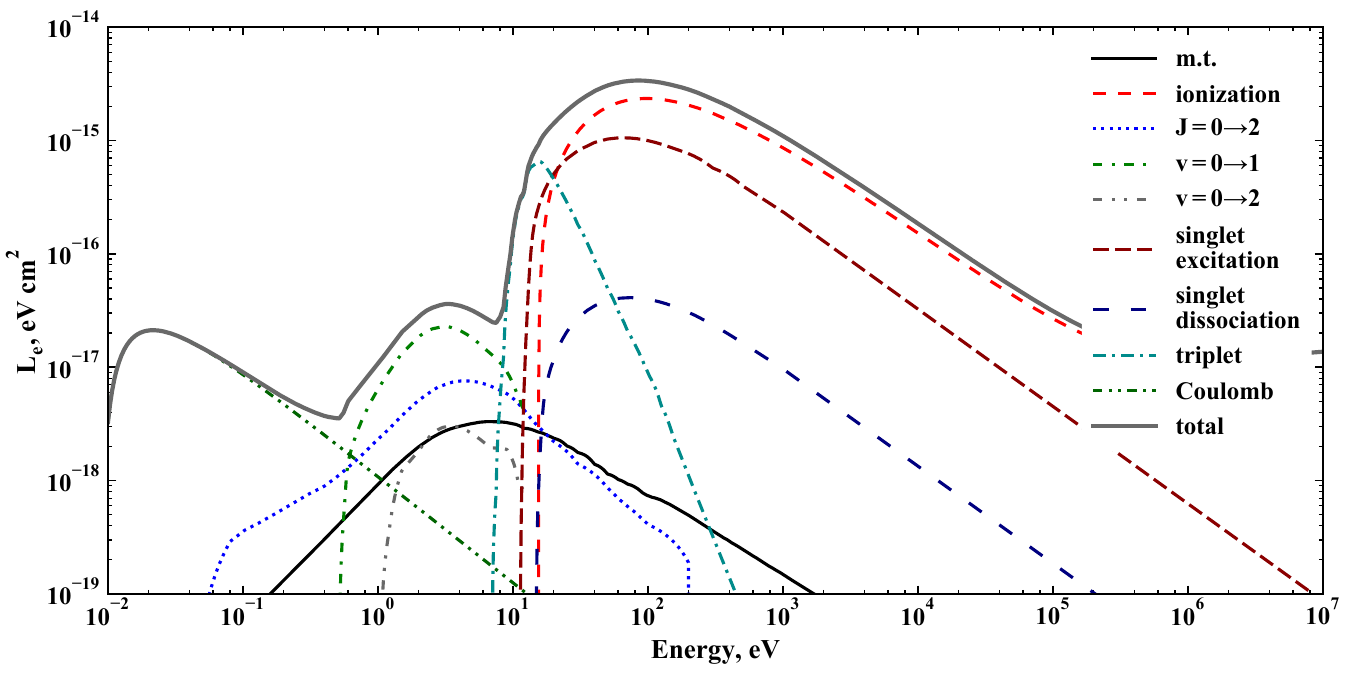}
\caption{Energy loss function for electrons in pure H$_2$ gas. Ionisation fraction of the gas $x_{e}$ is set equal to $10^{-7}$ in the calculations of Coulomb energy loss.}\label{fig_app_1}
\end{figure*}

The energy loss rate of a fast electron propagating through pure H$_2$ gas is:

\begin{equation}
\frac{d\varepsilon}{dt} = -L_{e}(\varepsilon) v_e(\varepsilon) n_{\rm H_2},
\end{equation}

\noindent
where $L_{e}(\varepsilon)$ is the energy loss function. The energy loss function is given by \citep{Dalgarno1999,Padovani2024}:

\begin{equation}
\begin{split}
L_{e}(\varepsilon) & = \varepsilon \frac{2m_e}{m_{\rm H_2}} \sigma_{\rm H_2, m.t.}(\varepsilon) + \frac{1}{v_{e}(\varepsilon) n_{\rm H_2}} \left\vert \frac{dE_{\rm c}}{dt} \right\vert + \\[5pt]
& + \sum_{u} \sigma_{um}(\varepsilon) \Delta\varepsilon_{um} + \int\limits_{0}^{(\varepsilon - I)/2} d\varepsilon^{\prime} \frac{d\sigma_{\rm ion, H_2}(\varepsilon, \varepsilon^{\prime})}{d\varepsilon^{\prime}} (I + \varepsilon^{\prime}),
\end{split}
\end{equation}

\noindent
where $m$ is initial energy level of H$_2$, and $u$ denotes the energy level or the dissociative excitation of H$_2$, $\Delta\varepsilon_{um}$ is energy lost by electron in collision. Here we assume that H$_2$ molecule is in the ground level $(v,J) = (0, 0)$. We consider pure rotational excitation $J = 0 \to 2$, the vibrational excitation $v = 0 \to 1$ and $v = 0 \to 2$ (each includes the transitions with $\Delta J = 0$ and 2), excitation to the singlet electronic states ($B$, $B^{\prime}$, $EF$, $C$, $D$), dissociative excitation to these singlet states, and excitation to the triplet electronic states ($a$, $b$, $c$, $d$, $e$, $g$, $h$, $i$, $j$). Figure~\ref{fig_app_1} shows the electron energy loss function $L_{e}(\varepsilon)$ for pure H$_2$ gas. The cross sections for ro-vibrational excitation of H$_2$ are available for electron energies $\varepsilon \leq 12$~eV. We do not extrapolate these cross sections at higher electron energies. At electron energies $\varepsilon > 12$~eV, the excitation to the electronic states and ionisation become much more important than ro-vibrational excitation.

\section{The simulations of the time evolution of electron energy distribution}\label{secA2}

The electron energy range higher than 1~eV is divided into intervals whose lengths follow a geometric progression. In simulations, the number $N$ of logarithmic intervals per decade of energy is chosen equal to 100. For electron energies $0-1$~eV, the energy intervals have the same length equal to 0.023~eV, that is equal to the length of the first energy interval higher than 1~eV. The electron energy interval with number $i$ is $[\varepsilon_i, \varepsilon_{i+1}]$, and $\varepsilon_{i+1/2}$ is the centre of the interval. The number density of electrons having the energy in the interval $i$ is $n_{i}$. 

Consider electron collisions leading to the excitation of H$_2$. In these collisions, electron loses the amount of energy equal to the excitation energy $\Delta \varepsilon$ of the H$_2$ transition. For simplicity, all electrons in the energy interval are considered to have energy $\varepsilon_{i+1/2}$. Only energy intervals that are entirely above the energy threshold are considered, $\varepsilon_{i} > \Delta \varepsilon$. The electron energy after collision is $\varepsilon^{\prime} = \varepsilon_{i + \frac{1}{2}} - \Delta \varepsilon$, and for some interval with number $k$:

\begin{equation}
\varepsilon_{k + \frac{1}{2}} < \varepsilon^{\prime} < \varepsilon_{k + \frac{3}{2}}.
\end{equation}

\noindent
where $\varepsilon_{k + \frac{3}{2}}$ is the centre of the energy interval $k+1$. We derive probabilities $\omega_k$ and $\omega_{k+1}$ that the scattered electron is injected into interval with number $k$ or $k+1$, respectively:

\begin{equation}
\omega_k = \frac{\varepsilon_{k + \frac{3}{2}} - \varepsilon^{\prime}}{\varepsilon_{k + \frac{3}{2}} - \varepsilon_{k + \frac{1}{2}}}, \quad \omega_{k+1} = 1 - \omega_k.
\end{equation} 

\noindent
The electron energy is conserved:

\begin{equation}
\varepsilon^{\prime} = \varepsilon_{k + \frac{1}{2}} \omega_{k} + \varepsilon_{k + \frac{3}{2}} \omega_{k + 1}.
\end{equation}

\noindent
The rates of excitation process are multiplied by probabilities $\omega_k$ or $\omega_{k+1}$ to determine the rate of electron transition from one energy interval to the other, e.g.:

\begin{equation}
\frac{dn_k}{dt} = \omega_k n_{{\rm H_2}, m} \sigma_{um}\big(\varepsilon_{i + \frac{1}{2}} \big) v_{i + \frac{1}{2}} n_{i},
\end{equation}

\noindent
where $v_{i + \frac{1}{2}}$ is the speed of electron with energy $\varepsilon_{i + \frac{1}{2}}$, $\sigma_{um}(\varepsilon)$ is the cross section for the transition $m \to u$ in H$_2$ molecule.

For ionisation collisions, the cross sections are calculated for energies of the primary electron $\varepsilon_{i + \frac{1}{2}}$ and ejected electron $\varepsilon_{m + \frac{1}{2}}$. The energy of the ejected electron satisfies:

\begin{equation}
\varepsilon_{m + \frac{1}{2}} < \frac{1}{2}\left(\varepsilon_{i + \frac{1}{2}} - I \right),
\end{equation}

\noindent
where we call the slower one of the two electrons after collision the ejected electron and the faster one the scattered electron \citep{Kim1994}. The energy of the scattered electron is 

\begin{equation}
\varepsilon^{\prime} = \varepsilon_{i + \frac{1}{2}} - \varepsilon_{m + \frac{1}{2}} - I.
\end{equation}

\noindent
The probabilities for the population of energy intervals for the scattered electron are calculated the same as for the excitation process.

Continuous energy losses (elastic scattering and Coulomb energy loss) are characterized by energy loss rate $dE/dt$ -- the amount of energy that a fast particle loses per unit time moving in the medium. The number of electrons that are ejected from and injected in the energy interval $i$ per unit time due to continuous energy losses is:

\begin{equation}
\frac{dn_i}{dt} = -\left\vert \frac{dE_{i}}{dt} \right\vert \frac{n_i}{\varepsilon_{i+1} - \varepsilon_{i}} + \left\vert \frac{dE_{i+1}}{dt} \right\vert \frac{n_{i+1}}{\varepsilon_{i+2} - \varepsilon_{i+1}},
\end{equation}

\noindent
where $dE_{i}/dt$ is calculated at $\varepsilon_{i}$.

We have conducted test simulations with the number of logarithmic intervals per decade of energy equal to $N = 500$. The difference between the calculated energy deposition parameters for two simulations with $N = 100$ and $N = 500$ is about or less than 1 per cent.

\section{Excitation numbers per ion pair for H$_2$ ro-vibrational levels}\label{secA3}

In this section, we provide excitation numbers per ion pair for ro-vibrational levels of H$_2$ molecule. We present the results for primary electron energies 30, 50, 100, 200, 500, 1000~eV, the helium abundance He/H$_2$ = 0.2, and the ionisation fraction $x_e = 0$. The results are presented separately for para-H$_2$ and ortho-H$_2$. Tables~\ref{app3_table1} and \ref{app3_table2} provide the excitation numbers per ion pair for para-H$_2$ molecule that initially resides in the energy level $v^{\prime} = 0$, $J^{\prime} = 0$. Table~\ref{app3_table1} presents the results for direct ro-vibrational excitation of H$_2$ energy levels, Table~\ref{app3_table2} -- for excitation to electronic states and subsequent radiative cascading. In Tables~\ref{app3_table3} and \ref{app3_table4}, we provide the excitation numbers per ion pair for ortho-H$_2$ molecule that initially resides in the energy level $v^{\prime} = 0$, $J^{\prime} = 1$. The final levels with $J^{\prime\prime} = J^{\prime}$, and $J^{\prime\prime} = J^{\prime} + 2$ are considered. The excitation yields for $J^{\prime\prime} = J^{\prime} + 4$ are about two orders of magnitude lower and are not shown. The results for pure rotational excitation with $v^{\prime\prime} = v^{\prime} = 0$ are also presented in Tables. The excitation yields for an arbitrary ortho-/para-H$_2$ ratio may be obtained by summing the results for para-H$_2$ and ortho-H$_2$, weighted by energy level populations. The estimate of ro-vibrational excitation yields at ionisation fraction $x_e > 0$ is given in the next section.

\begin{sidewaystable}
\caption{$J^{\prime} = 0$, direct excitation}\label{app3_table1}
\begin{tabular}{@{}lcccccccccccc@{}}
\toprule
& \multicolumn{2}{@{}c@{}}{E = 30 eV} & \multicolumn{2}{@{}c@{}}{E = 50 eV} &  \multicolumn{2}{@{}c@{}}{E = 100 eV} &  \multicolumn{2}{@{}c@{}}{E = 200 eV} &  \multicolumn{2}{@{}c@{}}{E = 500 eV} &  \multicolumn{2}{@{}c@{}}{E = 1000 eV} \\
$v^{\prime\prime}$ & $J^{\prime\prime}$ = 0 & $J^{\prime\prime}$ = 2 & $J^{\prime\prime}$ = 0 & $J^{\prime\prime}$ = 2 & $J^{\prime\prime}$ = 0 & $J^{\prime\prime}$ = 2 & $J^{\prime\prime}$ = 0 & $J^{\prime\prime}$ = 2 & $J^{\prime\prime}$ = 0 & $J^{\prime\prime}$ = 2 & $J^{\prime\prime}$ = 0 & $J^{\prime\prime}$ = 2\\
\midrule
0 & 0.0 & 83.0 & 0.0 & 52.8 & 0.0 & 37.5 & 0.0 & 31.4 & 0.0 & 28.1 & 0.0 & 27.1 \\ 
1 & 6.13 & 7.09 & 3.90 & 4.53 & 2.76 & 3.20 & 2.32 & 2.67 & 2.08 & 2.40 & 2.00 & 2.31 \\ 
2 & 0.320 & 0.440 & 0.205 & 0.281 & 0.144 & 0.197 & 0.120 & 0.165 & 0.107 & 0.147 & 0.103 & 0.142 \\ 
3 & 0.026 & 0.045 & 0.016 & 0.028 & 0.011 & 0.020 & 0.010 & 0.016 & 0.009 & 0.015 & 0.008 & 0.014 \\ 
4 & 0.003 & 0.007 & 0.002 & 0.004 & 0.001 & 0.003 & 0.001 & 0.002 & 0.001 & 0.002 & 0.001 & 0.002 \\ 
\botrule
\end{tabular}
\footnotetext{The results are shown for He/H$_2$ = 0.2 and $x_e = 0$.}
\end{sidewaystable}

\begin{sidewaystable}
\caption{$J^{\prime} = 0$, electronic excitation and cascading}\label{app3_table2}
\begin{tabular}{@{}lcccccccccccc@{}}
\toprule
& \multicolumn{2}{@{}c@{}}{E = 30 eV} & \multicolumn{2}{@{}c@{}}{E = 50 eV} &  \multicolumn{2}{@{}c@{}}{E = 100 eV} &  \multicolumn{2}{@{}c@{}}{E = 200 eV} &  \multicolumn{2}{@{}c@{}}{E = 500 eV} &  \multicolumn{2}{@{}c@{}}{E = 1000 eV} \\
$v^{\prime\prime}$ & $J^{\prime\prime}$ = 0 & $J^{\prime\prime}$ = 2 & $J^{\prime\prime}$ = 0 & $J^{\prime\prime}$ = 2 & $J^{\prime\prime}$ = 0 & $J^{\prime\prime}$ = 2 & $J^{\prime\prime}$ = 0 & $J^{\prime\prime}$ = 2 & $J^{\prime\prime}$ = 0 & $J^{\prime\prime}$ = 2 & $J^{\prime\prime}$ = 0 & $J^{\prime\prime}$ = 2\\
\midrule
    0 & 0.077 & 0.075 & 0.066 & 0.062 & 0.061 & 0.056 & 0.059 & 0.053 & 0.057 & 0.051 & 0.055 & 0.049 \\ 
    1 & 0.051 & 0.052 & 0.043 & 0.042 & 0.040 & 0.038 & 0.038 & 0.036 & 0.037 & 0.034 & 0.036 & 0.033 \\ 
    2 & 0.046 & 0.046 & 0.039 & 0.037 & 0.036 & 0.033 & 0.034 & 0.031 & 0.033 & 0.029 & 0.032 & 0.029 \\ 
    3 & 0.048 & 0.047 & 0.041 & 0.039 & 0.038 & 0.034 & 0.036 & 0.032 & 0.035 & 0.031 & 0.034 & 0.030 \\ 
    4 & 0.046 & 0.046 & 0.039 & 0.037 & 0.036 & 0.033 & 0.034 & 0.031 & 0.033 & 0.029 & 0.032 & 0.028 \\ 
    5 & 0.045 & 0.044 & 0.039 & 0.036 & 0.036 & 0.032 & 0.035 & 0.031 & 0.033 & 0.029 & 0.032 & 0.028 \\ 
    6 & 0.039 & 0.040 & 0.033 & 0.033 & 0.031 & 0.029 & 0.029 & 0.028 & 0.028 & 0.026 & 0.028 & 0.026 \\ 
    7 & 0.031 & 0.034 & 0.026 & 0.027 & 0.024 & 0.024 & 0.023 & 0.023 & 0.022 & 0.022 & 0.022 & 0.021 \\ 
    8 & 0.028 & 0.029 & 0.024 & 0.023 & 0.022 & 0.021 & 0.021 & 0.019 & 0.020 & 0.018 & 0.020 & 0.018 \\ 
    9 & 0.022 & 0.028 & 0.019 & 0.022 & 0.017 & 0.020 & 0.016 & 0.018 & 0.016 & 0.017 & 0.015 & 0.017 \\ 
   10 & 0.019 & 0.026 & 0.016 & 0.020 & 0.014 & 0.018 & 0.014 & 0.017 & 0.013 & 0.016 & 0.013 & 0.015 \\ 
   11 & 0.017 & 0.025 & 0.014 & 0.020 & 0.012 & 0.018 & 0.012 & 0.016 & 0.011 & 0.015 & 0.011 & 0.015 \\ 
   12 & 0.015 & 0.024 & 0.012 & 0.019 & 0.011 & 0.017 & 0.010 & 0.016 & 0.010 & 0.015 & 0.010 & 0.014 \\ 
   13 & 0.012 & 0.021 & 0.010 & 0.017 & 0.009 & 0.015 & 0.009 & 0.014 & 0.008 & 0.013 & 0.008 & 0.013 \\ 
   14 & 0.008 & 0.013 & 0.007 & 0.011 & 0.006 & 0.010 & 0.006 & 0.009 & 0.006 & 0.009 & 0.006 & 0.009 \\
\botrule
\end{tabular}
\footnotetext{The results are shown for He/H$_2$ = 0.2 and $x_e = 0$.}
\end{sidewaystable}

\begin{sidewaystable}
\caption{$J^{\prime} = 1$, direct excitation}\label{app3_table3}
\begin{tabular}{@{}lcccccccccccc@{}}
\toprule
& \multicolumn{2}{@{}c@{}}{E = 30 eV} & \multicolumn{2}{@{}c@{}}{E = 50 eV} &  \multicolumn{2}{@{}c@{}}{E = 100 eV} &  \multicolumn{2}{@{}c@{}}{E = 200 eV} &  \multicolumn{2}{@{}c@{}}{E = 500 eV} &  \multicolumn{2}{@{}c@{}}{E = 1000 eV} \\
$v^{\prime\prime}$ & $J^{\prime\prime}$ = 0 & $J^{\prime\prime}$ = 2 & $J^{\prime\prime}$ = 0 & $J^{\prime\prime}$ = 2 & $J^{\prime\prime}$ = 0 & $J^{\prime\prime}$ = 2 & $J^{\prime\prime}$ = 0 & $J^{\prime\prime}$ = 2 & $J^{\prime\prime}$ = 0 & $J^{\prime\prime}$ = 2 & $J^{\prime\prime}$ = 0 & $J^{\prime\prime}$ = 2\\
\midrule
0 & 0.0 & 48.3 & 0.0 & 30.7 & 0.0 & 21.8 & 0.0 & 18.2 & 0.0 & 16.3 & 0.0 & 15.7 \\ 
1 & 9.11 & 4.23 & 5.80 & 2.70 & 4.10 & 1.90 & 3.43 & 1.59 & 3.07 & 1.42 & 2.96 & 1.37 \\ 
2 & 0.503 & 0.264 & 0.322 & 0.168 & 0.226 & 0.118 & 0.188 & 0.098 & 0.168 & 0.088 & 0.162 & 0.085 \\ 
3 & 0.045 & 0.027 & 0.028 & 0.017 & 0.020 & 0.012 & 0.016 & 0.010 & 0.015 & 0.009 & 0.014 & 0.008 \\ 
4 & 0.006 & 0.004 & 0.004 & 0.002 & 0.002 & 0.002 & 0.002 & 0.001 & 0.002 & 0.001 & 0.002 & 0.001 \\ 
\botrule
\end{tabular}
\end{sidewaystable}

\begin{sidewaystable}
\caption{$J^{\prime} = 1$, electronic excitation and cascading}\label{app3_table4}
\begin{tabular}{@{}lcccccccccccc@{}}
\toprule
& \multicolumn{2}{@{}c@{}}{E = 30 eV} & \multicolumn{2}{@{}c@{}}{E = 50 eV} &  \multicolumn{2}{@{}c@{}}{E = 100 eV} &  \multicolumn{2}{@{}c@{}}{E = 200 eV} &  \multicolumn{2}{@{}c@{}}{E = 500 eV} &  \multicolumn{2}{@{}c@{}}{E = 1000 eV} \\
$v^{\prime\prime}$ & $J^{\prime\prime}$ = 0 & $J^{\prime\prime}$ = 2 & $J^{\prime\prime}$ = 0 & $J^{\prime\prime}$ = 2 & $J^{\prime\prime}$ = 0 & $J^{\prime\prime}$ = 2 & $J^{\prime\prime}$ = 0 & $J^{\prime\prime}$ = 2 & $J^{\prime\prime}$ = 0 & $J^{\prime\prime}$ = 2 & $J^{\prime\prime}$ = 0 & $J^{\prime\prime}$ = 2\\
\midrule
    0 & 0.108 & 0.047 & 0.092 & 0.039 & 0.085 & 0.035 & 0.081 & 0.034 & 0.078 & 0.032 & 0.076 & 0.031 \\ 
    1 & 0.072 & 0.032 & 0.060 & 0.026 & 0.055 & 0.023 & 0.053 & 0.022 & 0.050 & 0.021 & 0.049 & 0.020 \\ 
    2 & 0.066 & 0.028 & 0.055 & 0.023 & 0.050 & 0.020 & 0.048 & 0.019 & 0.046 & 0.018 & 0.044 & 0.018 \\ 
    3 & 0.066 & 0.029 & 0.055 & 0.024 & 0.051 & 0.021 & 0.048 & 0.020 & 0.046 & 0.019 & 0.045 & 0.018 \\ 
    4 & 0.066 & 0.027 & 0.055 & 0.022 & 0.050 & 0.019 & 0.048 & 0.018 & 0.046 & 0.017 & 0.045 & 0.017 \\ 
    5 & 0.063 & 0.027 & 0.053 & 0.022 & 0.049 & 0.019 & 0.047 & 0.018 & 0.045 & 0.017 & 0.043 & 0.017 \\ 
    6 & 0.057 & 0.023 & 0.048 & 0.019 & 0.044 & 0.016 & 0.042 & 0.015 & 0.041 & 0.015 & 0.040 & 0.014 \\ 
    7 & 0.049 & 0.020 & 0.041 & 0.016 & 0.038 & 0.014 & 0.036 & 0.014 & 0.035 & 0.013 & 0.034 & 0.012 \\ 
    8 & 0.040 & 0.019 & 0.034 & 0.015 & 0.031 & 0.013 & 0.030 & 0.012 & 0.029 & 0.012 & 0.028 & 0.011 \\ 
    9 & 0.035 & 0.017 & 0.029 & 0.013 & 0.026 & 0.012 & 0.025 & 0.011 & 0.024 & 0.010 & 0.024 & 0.010 \\ 
   10 & 0.030 & 0.016 & 0.025 & 0.012 & 0.023 & 0.011 & 0.022 & 0.010 & 0.021 & 0.009 & 0.020 & 0.009 \\ 
   11 & 0.027 & 0.015 & 0.022 & 0.012 & 0.020 & 0.010 & 0.019 & 0.010 & 0.018 & 0.009 & 0.018 & 0.009 \\ 
   12 & 0.025 & 0.015 & 0.021 & 0.012 & 0.018 & 0.010 & 0.017 & 0.009 & 0.016 & 0.009 & 0.016 & 0.008 \\ 
   13 & 0.021 & 0.013 & 0.018 & 0.010 & 0.016 & 0.009 & 0.015 & 0.008 & 0.014 & 0.008 & 0.014 & 0.008 \\ 
   14 & 0.014 & 0.008 & 0.011 & 0.007 & 0.010 & 0.006 & 0.010 & 0.006 & 0.009 & 0.006 & 0.009 & 0.006 \\ 
\botrule
\end{tabular}
\end{sidewaystable}

\section{Parameters of the electron energy degradation in H$_2$-He gas mixture}\label{secA4}
In this section, we present the results for electron energy deposition parameters for various primary electron energies and gas ionisation fractions. The helium abundance is He/H$_2$ = 0.2, the ortho-to-para-H$_2$ ratio is equal to 3. We provide the mean energy per ion pair, the number of excitations to H$_2$ electronic states B and C in Tables~\ref{app4_table1}, \ref{app4_table2}, and \ref{app4_table3}, respectively. These parameters have a low sensitivity to the gas ionisation fraction. Tables~\ref{app4_table4} and \ref{app4_table5} present the number of H$_2$ dissociations per ion pair and dissociation heat input, respectively. Table~\ref{app4_table6} gives the fraction of energy lost in ro-vibrational excitation of H$_2$ energy levels, including pure rotational excitation. Table~\ref{app4_table7} provides the fraction of energy lost in direct ro-vibrational excitation to the state $v=1$. Table~\ref{app4_table8} presents the heating efficiency.

The yields for direct ro-vibrational excitation of H$_2$ energy levels at ionisation fraction $x_e > 0$ can be estimated based on the data presented in Tables~\ref{app3_table1}, \ref{app3_table3} and \ref{app4_table7}. For example, from the data given in Table~\ref{app4_table7} follows that the fraction of primary energy lost in excitation to $v=1$ state at the gas ionisation fraction $x_e = 10^{-3}$ is approximately 50 times lower than at $x_e = 0$, at primary electron energy of 1~keV. Thus, the excitation yields from Tables~\ref{app3_table1} and \ref{app3_table3} must be divided by this value to estimate the excitation yields at $x_e = 10^{-3}$, where we take into account that the mean energy per ion pair has a weak dependence on the ionisation fraction. The yields for excitation to ro-vibrational levels through electronically excited states (Tables~\ref{app3_table2} and \ref{app3_table4}) have a weak dependence on the ionisation fraction. 

There is a negligible dependence on the initial energy level of H$_2$ for all energy degradation parameters except the heating efficiency. It is explained by the calculation method: the heating efficiency is mainly determined by the collisional de-excitation of the lowest rotational levels of H$_2$ molecule at low gas ionisation fraction. The collisional de-excitation competes with radiative de-excitation, and the heating efficiency is different for ortho- and para-H$_2$ at low gas density and low gas temperature.

\begin{table}[h]
\caption{Mean energy per ion pair}\label{app4_table1}
\begin{tabular}{@{}lccccccc@{}}
\toprule
Energy, eV & $x_e = 0$ & $10^{-7}$ & $10^{-6}$ & $10^{-5}$ & $10^{-4}$ & $10^{-3}$ & $10^{-2}$ \\
\midrule
    30 & 60.6 & 60.6 & 60.6 & 60.8 & 62.6 & 72.0 & 193.7 \\ 
    50 & 45.9 & 45.9 & 45.9 & 45.9 & 46.6 & 51.7 & 89.7 \\ 
   100 & 39.4 & 39.4 & 39.4 & 39.4 & 39.8 & 42.4 & 59.7 \\ 
   200 & 36.8 & 36.8 & 36.8 & 36.9 & 37.1 & 38.9 & 49.8 \\ 
   500 & 35.3 & 35.3 & 35.3 & 35.4 & 35.5 & 36.9 & 44.6 \\ 
  1000 & 34.7 & 34.7 & 34.7 & 34.8 & 34.9 & 36.1 & 42.8 \\ 
\botrule
\end{tabular}
\end{table}

\begin{table}[h]
\caption{Number of excitations to B state per H$_2^+$/H$^+$ ion}\label{app4_table2}
\begin{tabular}{@{}lccccccc@{}}
\toprule
Energy, eV & $x_e = 0$ & $10^{-7}$ & $10^{-6}$ & $10^{-5}$ & $10^{-4}$ & $10^{-3}$ & $10^{-2}$ \\
\midrule
    30 & 0.56 & 0.56 & 0.56 & 0.56 & 0.56 & 0.51 & 0.54 \\ 
    50 & 0.46 & 0.46 & 0.46 & 0.46 & 0.46 & 0.44 & 0.42 \\ 
   100 & 0.41 & 0.41 & 0.41 & 0.41 & 0.41 & 0.40 & 0.37 \\ 
   200 & 0.39 & 0.39 & 0.39 & 0.39 & 0.39 & 0.38 & 0.36 \\ 
   500 & 0.37 & 0.37 & 0.37 & 0.37 & 0.37 & 0.36 & 0.34 \\ 
  1000 & 0.36 & 0.36 & 0.36 & 0.36 & 0.36 & 0.35 & 0.34 \\ 
\botrule
\end{tabular}
\end{table}

\begin{table}[h]
\caption{Number of excitations to C state per H$_2^+$/H$^+$ ion}\label{app4_table3}
\begin{tabular}{@{}lccccccc@{}}
\toprule
Energy, eV & $x_e = 0$ & $10^{-7}$ & $10^{-6}$ & $10^{-5}$ & $10^{-4}$ & $10^{-3}$ & $10^{-2}$ \\
\midrule
    30 & 0.43 & 0.43 & 0.43 & 0.43 & 0.43 & 0.40 & 0.42 \\ 
    50 & 0.38 & 0.38 & 0.38 & 0.38 & 0.38 & 0.37 & 0.36 \\ 
   100 & 0.37 & 0.37 & 0.37 & 0.37 & 0.36 & 0.36 & 0.35 \\ 
   200 & 0.36 & 0.36 & 0.36 & 0.36 & 0.36 & 0.35 & 0.35 \\ 
   500 & 0.35 & 0.35 & 0.35 & 0.35 & 0.35 & 0.35 & 0.34 \\ 
  1000 & 0.35 & 0.35 & 0.35 & 0.35 & 0.35 & 0.34 & 0.34 \\ 
\botrule
\end{tabular}
\end{table}

\begin{table}[h]
\caption{Number of H$_2$ dissociations per H$_2^+$/H$^+$ ion}\label{app4_table4}
\begin{tabular}{@{}lccccccc@{}}
\toprule
Energy, eV & $x_e = 0$ & $10^{-7}$ & $10^{-6}$ & $10^{-5}$ & $10^{-4}$ & $10^{-3}$ & $10^{-2}$ \\
\midrule
    30 & 1.89 & 1.89 & 1.88 & 1.83 & 1.65 & 1.15 & 1.20 \\ 
    50 & 1.04 & 1.04 & 1.04 & 1.01 & 0.89 & 0.68 & 0.48 \\ 
   100 & 0.70 & 0.70 & 0.70 & 0.68 & 0.60 & 0.43 & 0.28 \\ 
   200 & 0.57 & 0.57 & 0.57 & 0.55 & 0.48 & 0.35 & 0.22 \\ 
   500 & 0.50 & 0.50 & 0.50 & 0.49 & 0.42 & 0.30 & 0.19 \\ 
  1000 & 0.48 & 0.48 & 0.48 & 0.46 & 0.41 & 0.29 & 0.18 \\ 
\botrule
\end{tabular}
\end{table}

\begin{table}[h]
\caption{Dissociation heat input}\label{app4_table5}
\begin{tabular}{@{}lccccccc@{}}
\toprule
Energy, eV & $x_e = 0$ & $10^{-7}$ & $10^{-6}$ & $10^{-5}$ & $10^{-4}$ & $10^{-3}$ & $10^{-2}$ \\ 
\midrule
    30 & 0.132 & 0.132 & 0.131 & 0.126 & 0.107 & 0.062 & 0.0234 \\ 
    50 & 0.090 & 0.090 & 0.089 & 0.086 & 0.072 & 0.045 & 0.0162 \\ 
   100 & 0.065 & 0.065 & 0.065 & 0.063 & 0.052 & 0.031 & 0.0107 \\ 
   200 & 0.054 & 0.054 & 0.054 & 0.051 & 0.042 & 0.024 & 0.0079 \\ 
   500 & 0.047 & 0.047 & 0.047 & 0.045 & 0.037 & 0.021 & 0.0065 \\ 
  1000 & 0.046 & 0.045 & 0.045 & 0.043 & 0.035 & 0.020 & 0.0061 \\ 
\botrule
\end{tabular}
\end{table}

\begin{table}[h]
\caption{Fraction of energy lost in ro-vibrational excitation of H$_2$}\label{app4_table6}
\begin{tabular}{@{}lccccccc@{}}
\toprule
Energy, eV & $x_e = 0$ & $10^{-7}$ & $10^{-6}$ & $10^{-5}$ & $10^{-4}$ & $10^{-3}$ & $10^{-2}$ \\
\midrule
    30 & 0.192 & 0.176 & 0.150 & 0.091 & 0.026 & 0.0054 & 0.00089 \\ 
    50 & 0.162 & 0.148 & 0.127 & 0.076 & 0.021 & 0.0040 & 0.00068 \\ 
   100 & 0.133 & 0.122 & 0.104 & 0.062 & 0.017 & 0.0031 & 0.00051 \\ 
   200 & 0.119 & 0.109 & 0.093 & 0.055 & 0.015 & 0.0027 & 0.00043 \\ 
   500 & 0.111 & 0.101 & 0.086 & 0.051 & 0.014 & 0.0024 & 0.00035 \\ 
  1000 & 0.109 & 0.099 & 0.084 & 0.050 & 0.014 & 0.0024 & 0.00033 \\ 
\botrule
\end{tabular}
\end{table}

\begin{table}[h]
\caption{Fraction of energy lost in excitation to the state $v=1$ of H$_2$}\label{app4_table7}
\begin{tabular}{@{}lccccccc@{}}
\toprule
Energy, eV & $x_e = 0$ & $10^{-7}$ & $10^{-6}$ & $10^{-5}$ & $10^{-4}$ & $10^{-3}$ & $10^{-2}$ \\
\midrule
    30 & 0.118 & 0.114 & 0.099 & 0.058 & 0.016 & 0.0028 & 0.00036 \\ 
    50 & 0.099 & 0.096 & 0.083 & 0.049 & 0.013 & 0.0021 & 0.00026 \\ 
   100 & 0.082 & 0.079 & 0.068 & 0.040 & 0.011 & 0.0017 & 0.00019 \\ 
   200 & 0.073 & 0.071 & 0.061 & 0.035 & 0.0093 & 0.0015 & 0.00017 \\ 
   500 & 0.068 & 0.066 & 0.057 & 0.033 & 0.0086 & 0.0013 & 0.00015 \\ 
  1000 & 0.067 & 0.065 & 0.056 & 0.032 & 0.0084 & 0.0013 & 0.00015 \\ 
\botrule
\end{tabular}
\end{table}

\begin{table}[h]
\caption{Heating efficiency}\label{app4_table8}
\begin{tabular}{@{}lccccccc@{}}
\toprule
Energy, eV & $x_e = 0$ & $10^{-7}$ & $10^{-6}$ & $10^{-5}$ & $10^{-4}$ & $10^{-3}$ & $10^{-2}$ \\
\midrule
    30 & 0.084 & 0.090 & 0.108 & 0.161 & 0.253 & 0.44 & 0.78 \\ 
    50 & 0.071 & 0.076 & 0.091 & 0.134 & 0.206 & 0.34 & 0.64 \\ 
   100 & 0.059 & 0.063 & 0.075 & 0.111 & 0.168 & 0.27 & 0.51 \\ 
   200 & 0.053 & 0.056 & 0.067 & 0.099 & 0.149 & 0.23 & 0.43 \\ 
   500 & 0.049 & 0.052 & 0.063 & 0.093 & 0.138 & 0.21 & 0.38 \\ 
  1000 & 0.048 & 0.051 & 0.062 & 0.091 & 0.135 & 0.20 & 0.36 \\ 
\botrule
\end{tabular}
\footnotetext{The results are shown for ortho-to-para-H$_2$ ratio equal to 3, hydrogen molecule number density $n_{\rm H_2} = 10^4$~cm$^{-3}$, and neutral gas kinetic temperature $T = 15$~K.}
\end{table}

\end{appendices}

\section*{Funding statement and Ethics approval}
Not Applicable.


\bibliography{h2_excitation_bib}

@ARTICLE{Abgrall2000,
       author = {{Abgrall}, H. and {Roueff}, E. and {Drira}, I.},
        title = "{Total transition probability and spontaneous radiative dissociation of B, C, B' and D states of molecular hydrogen}",
      journal = {\aaps},
         year = 2000,
        month = jan,
       volume = {141},
        pages = {297-300},
          doi = {10.1051/aas:2000121},
       adsurl = {https://ui.adsabs.harvard.edu/abs/2000A&AS..141..297A}
}

@BOOK{Avakyan1998,
   author = {{Avakyan}, S.~V. and {Il’in}, R.~N. and {Lavrov}, V.~M. and {Ogurtsov}, G.~N.},
    title = "{Collision Processes and Excitation of Ultraviolet Emission from Planetary Atmospheric Gases: A Handbook of Cross Sections}",
   address = {London},
   editor = {{Avakyan}, S.~V.},
publisher = {Gordon and Breach Publishing Group},
     year = {1998},
pagetotal = {356}
}

@ARTICLE{Bialy2026,
       author = {{Bialy}, Shmuel and {Chemke}, Amit and {Neufeld}, David A. and {Muzerolle Page}, James and {Ivlev}, Alexei V. and {Belli}, Sirio and {Gaches}, Brandt A.~L. and {Godard}, Benjamin and {Bisbas}, Thomas G. and {Caselli}, Paola and {Jacob}, Arshia M. and {Padovani}, Marco and {Rab}, Christian and {Silsbee}, Kedron and {Porter}, Troy A. and {Makarenko}, Ekaterina I.},
        title = "{Direct detection of cosmic-ray-excited H$_{2}$ in interstellar space}",
      journal = {Nature Astronomy},
         year = 2026,
       volume = {10},
        pages = {540--547},
        month = feb,
          doi = {10.1038/s41550-025-02771-9},
archivePrefix = {arXiv},
       eprint = {2508.20168},
 primaryClass = {astro-ph.GA},
       adsurl = {https://ui.adsabs.harvard.edu/abs/2026NatAs.tmp...28B}
}

@book{ICRU_31, 
	author={{Bichsel}, H. and {Peirson}, D.~H. and {Boring}, J.~W. and {Green}, A. and {Inokuti}, M. and {Hurst}, G.}, 
	title={ICRU report 31 Average energy required to produce an ion pair}, 
	publisher={Washington, DC: International Commission on Radiation Units and Measurements},
	year={1979}
}

@article{Buckman1990,
        title = "{Near-threshold vibrational excitation of H$_2$ by electron impact: Resolution of discrepancies between experiment and theory}",
       author = {Buckman, Stephen J. and Brunger, M. J. and Newman, D. S. and Snitchler, G. and Alston, S. and Norcross, D. W. and Morrison, Michael A. and Saha, B. C. and Danby, G. and Trail, W. K.},
      journal = {\prl},
       volume = {65},
        issue = {26},
        pages = {3253--3256},
     numpages = {0},
         year = {1990},
        month = {Dec},
    publisher = {American Physical Society},
          doi = {10.1103/PhysRevLett.65.3253}
}

@ARTICLE{Combecher1980,
       author = {{Combecher}, D.},
        title = "{Measurement of W Values of Low-Energy Electrons in Several Gases}",
      journal = {Radiation Research},
         year = 1980,
        month = nov,
       volume = {84},
       number = {2},
        pages = {189-218},
          doi = {10.2307/3575293},
       adsurl = {https://ui.adsabs.harvard.edu/abs/1980RadR...84..189C}
}

@ARTICLE{Cravens1975,
       author = {{Cravens}, T.~E. and {Victor}, G.~A. and {Dalgarno}, A.},
        title = "{The absorption of energetic electrons by molecular hydrogen gas}",
      journal = {\planss},
         year = 1975,
        month = jul,
       volume = {23},
       number = {7},
        pages = {1059-1070},
          doi = {10.1016/0032-0633(75)90196-8},
       adsurl = {https://ui.adsabs.harvard.edu/abs/1975P&SS...23.1059C}
}

@ARTICLE{Crompton1970,
       author = {{Crompton}, R.~W. and {Elford}, M.~T. and {Robertson}, A.~G.},
        title = "{The momentum transfer cross section for electrons in helium derived from drift velocities at 77{\textdegree}K}",
      journal = {Australian Journal of Physics},
         year = 1970,
        month = oct,
       volume = {23},
        pages = {667-682},
          doi = {10.1071/PH700667},
       adsurl = {https://ui.adsabs.harvard.edu/abs/1970AuJPh..23..667C}
}

@ARTICLE{Dalgarno1999,
       author = {{Dalgarno}, A. and {Yan}, Min and {Liu}, Weihong},
        title = "{Electron Energy Deposition in a Gas Mixture of Atomic and Molecular Hydrogen and Helium}",
      journal = {\apjs},
         year = 1999,
        month = nov,
       volume = {125},
       number = {1},
        pages = {237-256},
          doi = {10.1086/313267},
       adsurl = {https://ui.adsabs.harvard.edu/abs/1999ApJS..125..237D}
}

@BOOK{Draine2011,
   author = {{Draine}, B.~T.},
    title = "{Physics of the Interstellar and Intergalactic Medium}",
booktitle = {Physics of the Interstellar and Intergalactic Medium by Bruce T.~Draine.~Princeton University Press, 2011.~ISBN: 978-0-691-12214-4},
   address = {Princeton},
publisher = {Princeton University Press},
     year = 2011,
   adsurl = {http://adsabs.harvard.edu/abs/2011piim.book.....D}
}

@ARTICLE{Ehrhardt1968,
       author = {{Ehrhardt}, H. and {Langhans}, L. and {Linder}, F. and {Taylor}, H.~S.},
        title = "{Resonance Scattering of Slow Electrons from H$_{2}$ and CO Angular Distributions}",
      journal = {Physical Review},
         year = 1968,
        month = sep,
       volume = {173},
       number = {1},
        pages = {222-230},
          doi = {10.1103/PhysRev.173.222},
       adsurl = {https://ui.adsabs.harvard.edu/abs/1968PhRv..173..222E}
}

@ARTICLE{England1988,
       author = {{England}, J.~P. and {Elford}, M.~T. and {Crompton}, R.~W.},
        title = "{A study of the vibrational excitation of H$_{2}$ by measurements of the drift velocity of electrons in H$_{2}$-Ne mixtures}",
      journal = {Australian Journal of Physics},
         year = 1988,
        month = jan,
       volume = {41},
        pages = {573--586},
          doi = {10.1071/PH880573},
       adsurl = {https://ui.adsabs.harvard.edu/abs/1988AuJPh..41..573E}
}

@ARTICLE{Fantz2006,
       author = {{Fantz}, U. and {W{\"u}nderlich}, D.},
        title = "{Franck Condon factors, transition probabilities, and radiative lifetimes for hydrogen molecules and their isotopomeres}",
      journal = {Atomic Data and Nuclear Data Tables},
         year = 2006,
        month = nov,
       volume = {92},
       number = {6},
        pages = {853-973},
          doi = {10.1016/j.adt.2006.05.001},
       adsurl = {https://ui.adsabs.harvard.edu/abs/2006ADNDT..92..853F}
}

@ARTICLE{Ferland2017,
       author = {{Ferland}, G.~J. and {Chatzikos}, M. and {Guzm{\'a}n}, F. and {Lykins}, M.~L. and {van Hoof}, P.~A.~M. and {Williams}, R.~J.~R. and {Abel}, N.~P. and {Badnell}, N.~R. and {Keenan}, F.~P. and {Porter}, R.~L. and {Stancil}, P.~C.},
        title = "{The 2017 Release Cloudy}",
      journal = {\rmxaa},
         year = 2017,
        month = oct,
       volume = {53},
        pages = {385-438},
          doi = {10.48550/arXiv.1705.10877},
archivePrefix = {arXiv},
       eprint = {1705.10877},
 primaryClass = {astro-ph.GA},
       adsurl = {https://ui.adsabs.harvard.edu/abs/2017RMxAA..53..385F}
}

@ARTICLE{FlowerRoueff1998,
       author = {{Flower}, D.~R. and {Roueff}, E.},
        title = "{Rovibrational relaxation in collisions between H$_2$ molecules: I. Transitions induced by ground state para-H$_2$}",
      journal = {Journal of Physics B: Atomic, Molecular and Optical Physics},
         year = {1998},
        month = {jul},
       volume = {31},
       number = {13},
        pages = {2935-2947},
          doi = {10.1088/0953-4075/31/13/012}
}

@ARTICLE{FlowerRoueff1999,
       author = {{Flower}, D.~R. and {Roueff}, E.},
        title = "{Rovibrational relaxation in collisions between H$_{2}$ molecules: II. Influence of the rotational state of the perturber}",
      journal = {Journal of Physics B: Atomic, Molecular and Optical Physics},
         year = {1999},
        month = {jul},
       volume = {32},
       number = {14},
        pages = {3399-3407},
          doi = {10.1088/0953-4075/32/14/310}
}

@ARTICLE{FlowerRoueffZeippen1998,
       author = {{Flower}, D.~R. and {Roueff}, E. and {Zeippen}, C.~J.},
        title = "{Rovibrational excitation of H$_2$ molecules by He atoms}",
      journal = {Journal of Physics B: Atomic, Molecular and Optical Physics},
         year = {1998},
        month = {mar},
       volume = {31},
       number = {5},
        pages = {1105-1113},
          doi = {10.1088/0953-4075/31/5/017}
}

@article{Gardner2022,
  title     = "{Enabling new flexibility in the {SUNDIALS} suite of nonlinear and differential/algebraic equation solvers}",
  author    = {Gardner, David J. and Reynolds, Daniel R. and Woodward, Carol S. and Balos, Cody J.},
  journal   = {ACM Transactions on Mathematical Software (TOMS)},
  year      = {2022},
  publisher = {Association for Computing Machinery},
     volume = {48},
     number = {3},
      month = sep,
  articleno = {31},
   numpages = {24},
  doi       = {10.1145/3539801}
}

@ARTICLE{Genevriez2019,
       author = {{G{\'e}n{\'e}vriez}, Matthieu and {Defrance}, Pierre and {Jureta}, Jozo J. and {Urbain}, Xavier},
        title = "{Absolute total cross sections for electron-impact double ionization of He(1s2s $^{3}$S) and He$^{-}$(1s2s2p $^{4}$P)}",
      journal = {European Physical Journal D},
         year = 2019,
        month = feb,
       volume = {73},
       number = {2},
          eid = {30},
        pages = {30},
          doi = {10.1140/epjd/e2018-90566-y},
       adsurl = {https://ui.adsabs.harvard.edu/abs/2019EPJD...73...30G}
}

@ARTICLE{Glassgold1973,
       author = {{Glassgold}, A.~E. and {Langer}, William D.},
        title = "{Heating of Molecular-Hydrogen Clouds by Cosmic Rays and X-Rays}",
      journal = {\apj},
         year = 1973,
        month = dec,
       volume = {186},
        pages = {859-888},
          doi = {10.1086/152552},
       adsurl = {https://ui.adsabs.harvard.edu/abs/1973ApJ...186..859G}
}

@ARTICLE{Glassgold2012,
       author = {{Glassgold}, Alfred E. and {Galli}, Daniele and {Padovani}, Marco},
        title = "{Cosmic-Ray and X-Ray Heating of Interstellar Clouds and Protoplanetary Disks}",
      journal = {\apj},
         year = 2012,
        month = sep,
       volume = {756},
       number = {2},
          eid = {157},
        pages = {157},
          doi = {10.1088/0004-637X/756/2/157},
archivePrefix = {arXiv},
       eprint = {1208.0523},
 primaryClass = {astro-ph.GA},
       adsurl = {https://ui.adsabs.harvard.edu/abs/2012ApJ...756..157G}
}

@ARTICLE{Glass-Maujean1984,
       author = {{Glass-Maujean}, M. and {Quadrelli}, P. and {Dressler}, K.},
        title = "{Band Transition Moments Between Excited Singlet States of the H$_{2}$ Molecule, Nonadiabatic Eigenvectors, and Probabilities for Spontaneous Emission}",
      journal = {Atomic Data and Nuclear Data Tables},
         year = 1984,
        month = jan,
       volume = {30},
        pages = {273--300},
          doi = {10.1016/0092-640X(84)90003-2},
       adsurl = {https://ui.adsabs.harvard.edu/abs/1984ADNDT..30..273G}
}

@ARTICLE{Glass-Maujean2007,
       author = {{Glass-Maujean}, M. and {Klumpp}, S. and {Werner}, L. and {Ehresmann}, A. and {Schmoranzer}, H.},
        title = "{Study of the B$^{''}$B $^{1}${\ensuremath{\Sigma}}$_{u}$$^{+}$ state of H$_{2}$: Transition probabilities from the ground state, dissociative widths, and Fano parameters}",
      journal = {\jcp},
         year = 2007,
        month = apr,
       volume = {126},
       number = {14},
        pages = {144303},
          doi = {10.1063/1.2715928},
       adsurl = {https://ui.adsabs.harvard.edu/abs/2007JChPh.126n4303G}
}

@ARTICLE{Glass-Maujean2007b,
       author = {{Glass-Maujean}, M. and {Klumpp}, S. and {Werner}, L. and {Ehresmann}, A. and {Schmoranzer}, H.},
        title = "{Transition probabilities from the ground state of the $np\pi {}^{1}\Pi^{-}_{u}$ states of H$_{2}$}",
      journal = {Molecular Physics},
         year = 2007,
        month = jun,
       volume = {105},
       number = {11-12},
        pages = {1535-1542},
          doi = {10.1080/00268970701271935},
       adsurl = {https://ui.adsabs.harvard.edu/abs/2007MolPh.105.1535G}
}

@ARTICLE{Glass-Maujean2008,
       author = {{Glass-Maujean}, M. and {Klumpp}, S. and {Werner}, L. and {Ehresmann}, A. and {Schmoranzer}, H.},
        title = "{The study of the D$^{'}$ $^{1}${\ensuremath{\Pi}}$_{u}$ state of H$_{2}$: Transition probabilities from the ground state, predissociation yields, and natural linewidths}",
      journal = {\jcp},
         year = 2008,
        month = mar,
       volume = {128},
       number = {9},
        pages = {094312},
          doi = {10.1063/1.2835006},
       adsurl = {https://ui.adsabs.harvard.edu/abs/2008JChPh.128i4312G}
}

@ARTICLE{Glass-Maujean2009,
       author = {{Glass-Maujean}, M. and {Jungen}, Ch.},
        title = "{Nonadiabatic Ab Initio Multichannel Quantum Defect Theory Applied to Absolute Experimental Absorption Intensities in H$_2$}",
      journal = {Journal of Physical Chemistry A},
         year = 2009,
        month = jun,
       volume = {113},
       number = {47},
        pages = {13124-13132},
          doi = {10.1021/jp902846c},
       adsurl = {https://ui.adsabs.harvard.edu/abs/2009JPCA..11313124G}
}

@ARTICLE{Goldsmith2005,
       author = {{Goldsmith}, P.~F. and {Li}, D.},
        title = "{H I Narrow Self-Absorption in Dark Clouds: Correlations with Molecular Gas and Implications for Cloud Evolution and Star Formation}",
      journal = {\apj},
         year = 2005,
        month = apr,
       volume = {622},
       number = {2},
        pages = {938-958},
          doi = {10.1086/428032},
archivePrefix = {arXiv},
       eprint = {astro-ph/0412427},
 primaryClass = {astro-ph}
}

@ARTICLE{Gredel1995,
       author = {{Gredel}, R. and {Dalgarno}, A.},
        title = "{Infrared Response of H$_2$ to X-Rays}",
      journal = {\apj},
         year = 1995,
        month = jun,
       volume = {446},
        pages = {852--859},
          doi = {10.1086/175843},
       adsurl = {https://ui.adsabs.harvard.edu/abs/1995ApJ...446..852G}
}

@article{Hindmarsh2005,
  title     = "{{SUNDIALS}: Suite of nonlinear and differential/algebraic equation solvers}",
  author    = {Hindmarsh, Alan C and Brown, Peter N and Grant, Keith E and Lee, Steven L and Serban, Radu and Shumaker, Dan E and Woodward, Carol S},
  journal   = {ACM Transactions on Mathematical Software (TOMS)},
  publisher = {ACM},
  volume    = {31},
  number    = {3},
  pages     = {363--396},
  year      = {2005},
  doi       = {10.1145/1089014.1089020}
}

@ARTICLE{Horton2021,
       author = {{Horton}, Reese K. and {Scarlett}, Liam H. and {Zammit}, Mark C. and {Bray}, Igor and {Fursa}, Dmitry V.},
        title = "{Electron energy deposition in molecular hydrogen gas: a Monte Carlo simulation using convergent close-coupling cross sections}",
      journal = {Plasma Sources Science Technology},
         year = 2021,
        month = nov,
       volume = {30},
       number = {11},
          eid = {115004},
        pages = {115004},
          doi = {10.1088/1361-6595/ac27ba},
       adsurl = {https://ui.adsabs.harvard.edu/abs/2021PSST...30k5004H}
}

@article{Jesse1955,
  title = "{Ionization in Pure Gases and the Average Energy to Make an Ion Pair for Alpha and Beta Particles}",
  author = {{Jesse}, W. P. and {Sadauskis}, J.},
  journal = {Phys. Rev.},
  volume = {97},
  issue = {6},
  pages = {1668--1670},
  numpages = {0},
  year = {1955},
  month = {Mar},
  publisher = {American Physical Society},
  doi = {10.1103/PhysRev.97.1668}
}

@ARTICLE{Jones1973,
       author = {{Jones}, W.~M.},
        title = "{Some calculated quantities in the radiation chemistry of molecular hydrogen: Average energy per ion pair and numbers of singlet and triplet excitations per ion pair}",
      journal = {\jcp},
         year = 1973,
        month = nov,
       volume = {59},
       number = {10},
        pages = {5688-5695},
          doi = {10.1063/1.1679921},
       adsurl = {https://ui.adsabs.harvard.edu/abs/1973JChPh..59.5688J}
}

@ARTICLE{Kim1994,
       author = {{Kim}, Yong-Ki and {Rudd}, M. Eugene},
        title = "{Binary-encounter-dipole model for electron-impact ionization}",
      journal = {\pra},
         year = 1994,
        month = nov,
       volume = {50},
       number = {5},
        pages = {3954-3967},
          doi = {10.1103/PhysRevA.50.3954},
       adsurl = {https://ui.adsabs.harvard.edu/abs/1994PhRvA..50.3954K}
}

@ARTICLE{Kim2000,
       author = {{Kim}, Yong-Ki and {Santos}, Jos{\'e} Paulo and {Parente}, Fernando},
        title = "{Extension of the binary-encounter-dipole model to relativistic incident electrons}",
      journal = {\pra},
         year = 2000,
        month = nov,
       volume = {62},
       number = {5},
          eid = {052710},
        pages = {052710},
          doi = {10.1103/PhysRevA.62.052710},
       adsurl = {https://ui.adsabs.harvard.edu/abs/2000PhRvA..62e2710K}
}

@ARTICLE{Kossmann1990,
       author = {{Kossmann}, H. and {Schwarzkopf}, O. and {Schmidt}, V.},
        title = "{Absolute ionisation cross sections for electron impact on H$_{2}$}",
      journal = {Journal of Physics B: Atomic, Molecular and Optical Physics},
         year = 1990,
        month = jan,
       volume = {23},
       number = {2},
        pages = {301-313},
          doi = {10.1088/0953-4075/23/2/012}
}

@ARTICLE{Lique2015,
   author = {{Lique}, F.},
    title = "{Revisited study of the ro-vibrational excitation of H$_{2}$ by H: towards a revision of the cooling of astrophysical media}",
  journal = {\mnras},
     year = 2015,
    month = oct,
   volume = 453,
    pages = {810-818},
      doi = {10.1093/mnras/stv1683},
   adsurl = {http://adsabs.harvard.edu/abs/2015MNRAS.453..810L}
}

@ARTICLE{Liu2010,
       author = {{Liu}, Xianming and {Johnson}, Paul V. and {Malone}, Charles P. and {Young}, Jason A. and {Kanik}, Isik and {Shemansky}, Donald E.},
        title = "{Kinetic Energy Distribution of H(1s) from H$_{2}$ X $^{1}${\ensuremath{\Sigma}}$^{+}_{g}$ -- a $^{3}${\ensuremath{\Sigma}}$^{+}_{g}$ Excitation and Lifetimes and Transition Probabilities of a $^{3}${\ensuremath{\Sigma}}$^{+}_{g}$(v, J)}",
      journal = {\apj},
         year = 2010,
        month = jun,
       volume = {716},
       number = {1},
        pages = {701-711},
          doi = {10.1088/0004-637X/716/1/701},
archivePrefix = {arXiv},
       eprint = {0910.5025},
 primaryClass = {astro-ph.CO},
       adsurl = {https://ui.adsabs.harvard.edu/abs/2010ApJ...716..701L}
}

@ARTICLE{Miles1972,
       author = {{Miles}, W.~T. and {Thompson}, R. and {Green}, A.~E.~S.},
        title = "{Electron-Impact Cross Sections and Energy Deposition in Molecular Hydrogen}",
      journal = {Journal of Applied Physics},
         year = 1972,
        month = feb,
       volume = {43},
       number = {2},
        pages = {678-686},
          doi = {10.1063/1.1661176},
       adsurl = {https://ui.adsabs.harvard.edu/abs/1972JAP....43..678M}
}

@ARTICLE{Milloy1977,
       author = {{Milloy}, H.~B. and {Crompton}, R.~W.},
        title = "{Momentum-transfer cross section for electron-helium collisions in the range 4--12 eV}",
      journal = {\pra},
         year = 1977,
        month = may,
       volume = {15},
       number = {5},
        pages = {1847-1850},
          doi = {10.1103/PhysRevA.15.1847},
       adsurl = {https://ui.adsabs.harvard.edu/abs/1977PhRvA..15.1847M}
}

@ARTICLE{Nesterenok2024,
       author = {{Nesterenok}, A.~V.},
        title = "{Passage of a Gamma-Ray Burst through a Molecular Cloud: Cloud Ionization Structure}",
      journal = {Astronomy Letters},
         year = 2024,
        month = may,
       volume = {50},
       number = {2},
        pages = {99-119},
          doi = {10.1134/S1063773724700014},
archivePrefix = {arXiv},
       eprint = {2405.10730},
 primaryClass = {astro-ph.HE},
       adsurl = {https://ui.adsabs.harvard.edu/abs/2024AstL...50...99N}
}

@ARTICLE{Nesterenok2024b,
       author = {{Nesterenok}, A.~V.},
        title = "{Passage of a Gamma-Ray Burst through a Molecular Cloud: The Absorption of Its Afterglow in the X-ray Wavelength Range}",
      journal = {Astronomy Letters},
         year = 2024,
        month = aug,
       volume = {50},
       number = {8},
        pages = {510-522},
          doi = {10.1134/S1063773724700403},
archivePrefix = {arXiv},
       eprint = {2412.10021},
 primaryClass = {astro-ph.HE},
       adsurl = {https://ui.adsabs.harvard.edu/abs/2024AstL...50..510N}
}

@ARTICLE{Olivero1973,
       author = {{Olivero}, J.~J. and {Bass}, J.~N. and {Green}, A.~E.~S.},
        title = "{Photoelectron excitation of the Jupiter dayglow}",
      journal = {\jgr},
         year = 1973,
        month = jan,
       volume = {78},
       number = {16},
        pages = {2812--2826},
          doi = {10.1029/JA078i016p02812},
       adsurl = {https://ui.adsabs.harvard.edu/abs/1973JGR....78.2812O}
}

@ARTICLE{Padovani2022,
       author = {{Padovani}, Marco and {Bialy}, Shmuel and {Galli}, Daniele and {Ivlev}, Alexei V. and {Grassi}, Tommaso and {Scarlett}, Liam H. and {Rehill}, Una S. and {Zammit}, Mark C. and {Fursa}, Dmitry V. and {Bray}, Igor},
        title = "{Cosmic rays in molecular clouds probed by H$_{2}$ rovibrational lines. Perspectives for the James Webb Space Telescope}",
      journal = {\aap},
         year = 2022,
        month = feb,
       volume = {658},
          eid = {A189},
        pages = {A189},
          doi = {10.1051/0004-6361/202142560},
archivePrefix = {arXiv},
       eprint = {2201.08457},
 primaryClass = {astro-ph.GA},
       adsurl = {https://ui.adsabs.harvard.edu/abs/2022A&A...658A.189P}
}

@ARTICLE{Padovani2024,
       author = {{Padovani}, Marco and {Galli}, Daniele and {Scarlett}, Liam H. and {Grassi}, Tommaso and {Rehill}, Una S. and {Zammit}, Mark C. and {Bray}, Igor and {Fursa}, Dmitry V.},
        title = "{Ultraviolet H$_{2}$ luminescence in molecular clouds induced by cosmic rays}",
      journal = {\aap},
         year = 2024,
        month = feb,
       volume = {682},
          eid = {A131},
        pages = {A131},
          doi = {10.1051/0004-6361/202348168},
archivePrefix = {arXiv},
       eprint = {2312.02062},
 primaryClass = {astro-ph.GA},
       adsurl = {https://ui.adsabs.harvard.edu/abs/2024A&A...682A.131P}
}

@ARTICLE{Pagani2009,
       author = {{Pagani}, L. and {Vastel}, C. and {Hugo}, E. and {Kokoouline}, V. and {Greene}, C.~H. and {Bacmann}, A. and {Bayet}, E. and {Ceccarelli}, C. and {Peng}, R. and {Schlemmer}, S.},
        title = "{Chemical modeling of L183 (L134N): an estimate of the ortho/para H$_2$ ratio}",
      journal = {\aap},
         year = 2009,
        month = feb,
       volume = {494},
       number = {2},
        pages = {623-636},
          doi = {10.1051/0004-6361:200810587},
archivePrefix = {arXiv},
       eprint = {0810.1861},
 primaryClass = {astro-ph},
       adsurl = {https://ui.adsabs.harvard.edu/abs/2009A&A...494..623P}
}

@article{Plowman2026,
    doi = {10.1088/1361-6587/ae464e},
    year = {2026},
    month = {mar},
    publisher = {IOP Publishing},
    volume = {68},
    number = {3},
    pages = {035006},
    author = {Plowman, C T and Scarlett, L H and Zammit, M C and Bray, I and Fursa, D V},
    title = {Vibrationally resolved cross sections for proton collisions with hydrogen molecules},
    journal = {Plasma Physics and Controlled Fusion}
}

@ARTICLE{Ralchenko2008,
       author = {{Ralchenko}, Yu. and {Janev}, R.~K. and {Kato}, T. and {Fursa}, D.~V. and {Bray}, I. and {de Heer}, F.~J.},
        title = "{Electron-impact excitation and ionization cross sections for ground state and excited helium atoms}",
      journal = {Atomic Data and Nuclear Data Tables},
         year = 2008,
        month = jul,
       volume = {94},
       number = {4},
        pages = {603-622},
          doi = {10.1016/j.adt.2007.11.003},
       adsurl = {https://ui.adsabs.harvard.edu/abs/2008ADNDT..94..603R}
}

@ARTICLE{Rescigno1993,
       author = {{Rescigno}, T.~N. and {Elza}, B.~K. and {Lengsfield}, III, B.~H.},
        title = "{An ab initio treatment of near-threshold vibrational excitation of H$_{2}$ by electron impact: new perspectives on discrepancies between crossed-beam and swarm data}",
      journal = {Journal of Physics B: Atomic, Molecular and Optical Physics},
         year = {1993},
        month = {sep},
       volume = {26},
       number = {17},
        pages = {L567-L573},
          doi = {10.1088/0953-4075/26/17/007}
}

@ARTICLE{Rochman2014,
       author = {{Rochman}, D. and {van der Marck}, S.~C. and {Koning}, A.~J. and {Sj{\"o}strand}, H. and {Zwermann}, W.},
        title = "{Uncertainty Propagation with Fast Monte Carlo Techniques}",
      journal = {Nuclear Data Sheets},
         year = 2014,
        month = apr,
       volume = {118},
        pages = {367-369},
          doi = {10.1016/j.nds.2014.04.082},
       adsurl = {https://ui.adsabs.harvard.edu/abs/2014NDS...118..367R}
}

@ARTICLE{Roueff2019,
       author = {{Roueff}, E. and {Abgrall}, H. and {Czachorowski}, P. and {Pachucki}, K. and {Puchalski}, M. and {Komasa}, J.},
        title = "{The full infrared spectrum of molecular hydrogen}",
      journal = {\aap},
         year = 2019,
        month = oct,
       volume = {630},
          eid = {A58},
        pages = {A58},
          doi = {10.1051/0004-6361/201936249},
archivePrefix = {arXiv},
       eprint = {1909.11585},
 primaryClass = {physics.atom-ph},
       adsurl = {https://ui.adsabs.harvard.edu/abs/2019A&A...630A..58R}
}

@ARTICLE{Scarlett2021,
       author = {{Scarlett}, Liam H. and {Fursa}, Dmitry V. and {Zammit}, Mark C. and {Bray}, Igor and {Ralchenko}, Yuri and {Davie}, Kayla D.},
        title = "{Complete collision data set for electrons scattering on molecular hydrogen and its isotopologues: I. Fully vibrationally-resolved electronic excitation of H$_{2}$(X$^{1}$ {\ensuremath{\Sigma}}$_{g}$$^{+}$)}",
      journal = {Atomic Data and Nuclear Data Tables},
         year = 2021,
        month = jan,
       volume = {137},
          eid = {101361},
        pages = {101361},
          doi = {10.1016/j.adt.2020.101361},
       adsurl = {https://ui.adsabs.harvard.edu/abs/2021ADNDT.13701361S}
}

@ARTICLE{Scarlett2023,
	title = "{Elastic scattering and rotational excitation of H$_2$ by electron impact: Convergent close-coupling calculations}",
	author = {Scarlett, Liam H. and Rehill, Una S. and Zammit, Mark C. and Mori, Nicolas A. and Bray, Igor and Fursa, Dmitry V.},
	journal = {\pra},
	volume = {107},
	issue = {6},
	pages = {062804},
	numpages = {14},
	year = {2023},
	month = {Jun},
	publisher = {American Physical Society},
	doi = {10.1103/PhysRevA.107.062804},
}

@ARTICLE{Shemansky1985,
       author = {{Shemansky}, D.~E. and {Ajello}, J.~M. and {Hall}, D.~T.},
        title = "{Electron Impact Excitation of H$_2$: Rydberg Band Systems and the Benchmark Dissociative Cross Section of H Lyman-Alpha}",
      journal = {\apj},
         year = 1985,
        month = sep,
       volume = {296},
        pages = {765--773},
          doi = {10.1086/163493},
       adsurl = {https://ui.adsabs.harvard.edu/abs/1985ApJ...296..765S}
}

@ARTICLE{Spencer1954,
  title = "{Energy Spectrum Resulting from Electron Slowing Down}",
  author = {Spencer, L. V. and Fano, U.},
  journal = {Phys. Rev.},
  volume = {93},
  issue = {6},
  pages = {1172--1181},
  numpages = {0},
  year = {1954},
  month = {Mar},
  publisher = {American Physical Society},
  doi = {10.1103/PhysRev.93.1172}
}

@ARTICLE{Straub1996,
       author = {{Straub}, H.~C. and {Renault}, P. and {Lindsay}, B.~G. and {Smith}, K.~A. and {Stebbings}, R.~F.},
        title = "{Absolute partial cross sections for electron-impact ionization of H$_{2}$, N$_{2}$, and O$_{2}$ from threshold to 1000 eV}",
      journal = {\pra},
         year = 1996,
        month = sep,
       volume = {54},
       number = {3},
        pages = {2146-2153},
          doi = {10.1103/PhysRevA.54.2146},
       adsurl = {https://ui.adsabs.harvard.edu/abs/1996PhRvA..54.2146S}
}

@ARTICLE{Swartz1971,
       author = {{Swartz}, Wesley E. and {Nisbet}, John S. and {Green}, Alex E.~S.},
        title = "{Analytic expression for the energy-transfer rate from photoelectrons to thermal-electrons}",
      journal = {\jgr},
         year = 1971,
        month = dec,
       volume = {76},
       number = {34},
        pages = {8425-8426},
          doi = {10.1029/JA076i034p08425},
       adsurl = {https://ui.adsabs.harvard.edu/abs/1971JGR....76.8425S}
}

@ARTICLE{Tine1997,
       author = {{Tin{\'e}}, S. and {Lepp}, S. and {Gredel}, R. and {Dalgarno}, A.},
        title = "{Infrared Response of H$_{2}$ to X-Rays in Dense Clouds}",
      journal = {\apj},
         year = 1997,
        month = may,
       volume = {481},
       number = {1},
        pages = {282-295},
          doi = {10.1086/304048},
       adsurl = {https://ui.adsabs.harvard.edu/abs/1997ApJ...481..282T}
}

@ARTICLE{Trevisan2002,
       author = {{Trevisan}, Cynthia S. and {Tennyson}, Jonathan},
        title = "{Calculated rates for the electron impact dissociation of molecular hydrogen, deuterium and tritium}",
      journal = {Plasma Physics and Controlled Fusion},
         year = 2002,
        month = jul,
       volume = {44},
       number = {7},
        pages = {1263-1276},
          doi = {10.1088/0741-3335/44/7/315},
       adsurl = {https://ui.adsabs.harvard.edu/abs/2002PPCF...44.1263T}
}

@ARTICLE{Voit1991,
       author = {{Voit}, G.~M.},
        title = "{Energy Deposition by X-Ray Photoelectrons into Interstellar Molecular Clouds}",
      journal = {\apj},
         year = 1991,
        month = aug,
       volume = {377},
        pages = {158},
          doi = {10.1086/170344},
       adsurl = {https://ui.adsabs.harvard.edu/abs/1991ApJ...377..158V}
}

@ARTICLE{Wan2018,
       author = {{Wan}, Yier and {Yang}, B.~H. and {Stancil}, P.~C. and {Balakrishnan}, N. and {Parekh}, Nikhil J. and {Forrey}, R.~C.},
        title = "{Collisional Quenching of Highly Excited H$_{2}$ due to H$_{2}$ Collisions}",
      journal = {\apj},
         year = 2018,
        month = aug,
       volume = {862},
       number = {2},
          eid = {132},
        pages = {132},
          doi = {10.3847/1538-4357/aaccf8},
       adsurl = {https://ui.adsabs.harvard.edu/abs/2018ApJ...862..132W}
}

@ARTICLE{Weiss1956,
  title = "{Energy Required to Produce One Ion Pair in Several Noble Gases}",
  author = {{Weiss}, J. and {Bernstein}, W.},
  journal = {Phys. Rev.},
  volume = {103},
  issue = {5},
  pages = {1253},
  numpages = {0},
  year = {1956},
  month = {Sep},
  publisher = {American Physical Society},
  doi = {10.1103/PhysRev.103.1253}
}

@ARTICLE{Wolniewicz2007,
       author = {{Wolniewicz}, L.},
        title = "{Non-adiabatic energies of the a $^{3}\Sigma^{+}_{g}$ state of the hydrogen molecule}",
      journal = {Molecular Physics},
         year = 2007,
        month = jun,
       volume = {105},
       number = {11-12},
        pages = {1497-1503},
          doi = {10.1080/00268970701257082},
       adsurl = {https://ui.adsabs.harvard.edu/abs/2007MolPh.105.1497W}
}

@ARTICLE{Wunderlich2021,
       author = {{W{\"u}nderlich}, D.},
        title = "{Vibrationally resolved ionization cross sections for the ground state and electronically excited states of the hydrogen molecule and its isotopomeres}",
      journal = {Atomic Data and Nuclear Data Tables},
         year = 2021,
        month = jul,
       volume = {140},
          eid = {101424},
        pages = {101424},
          doi = {10.1016/j.adt.2021.101424},
       adsurl = {https://ui.adsabs.harvard.edu/abs/2021ADNDT.14001424W}
}

@ARTICLE{Xu1991,
       author = {{Xu}, Yueming and {McCray}, Richard},
        title = "{Energy Degradation of Fast Electrons in Hydrogen Gas}",
      journal = {\apj},
         year = 1991,
        month = jul,
       volume = {375},
        pages = {190--201},
          doi = {10.1086/170180},
       adsurl = {https://ui.adsabs.harvard.edu/abs/1991ApJ...375..190X}
}

@ARTICLE{Yoon2008,
       author = {{Yoon}, Jung-Sik and {Song}, Mi-Young and {Han}, Jeong-Min and {Hwang}, Sung Ha and {Chang}, Won-Seok and {Lee}, BongJu and {Itikawa}, Yukikazu},
        title = "{Cross Sections for Electron Collisions with Hydrogen Molecules}",
      journal = {Journal of Physical and Chemical Reference Data},
         year = 2008,
        month = jun,
       volume = {37},
       number = {2},
        pages = {913-931},
          doi = {10.1063/1.2838023},
       adsurl = {https://ui.adsabs.harvard.edu/abs/2008JPCRD..37..913Y}
}

@ARTICLE{Zammit2017a,
       author = {{Zammit}, Mark C. and {Fursa}, Dmitry V. and {Savage}, Jeremy S. and {Bray}, Igor},
        title = "{Electron- and positron-molecule scattering: development of the molecular convergent close-coupling method}",
      journal = {Journal of Physics B: Atomic, Molecular and Optical Physics},
         year = 2017,
        month = {may},
       volume = {50},
       number = {12},
        pages = {123001},
          doi = {10.1088/1361-6455/aa6e74}
}

@ARTICLE{Zammit2017b,
       author = {{Zammit}, Mark C. and {Savage}, Jeremy S. and {Fursa}, Dmitry V. and {Bray}, Igor},
        title = "{Electron-impact excitation of molecular hydrogen}",
      journal = {\pra},
         year = 2017,
        month = feb,
       volume = {95},
       number = {2},
          eid = {022708},
        pages = {022708},
          doi = {10.1103/PhysRevA.95.022708},
       adsurl = {https://ui.adsabs.harvard.edu/abs/2017PhRvA..95b2708Z}
}

@online{NIST,
  author = {{Kramida}, A. and {Ralchenko}, Yu. and {Reader}, J. and {NIST ASD Team (2024)}},
  title = {{NIST} Atomic Spectra Database (version 5.12)},
  year = 2024,
  url = {https://physics.nist.gov/asd},
  urldate = {2025-02-03}
}

\end{document}